\documentclass[twocolumn,
showpacs,
preprintnumbers,
amsmath,
amssymb,
amsfonts,
a4paper,
aps,
floatfix,
citeautoscript,
footinbib,
pra,
superscriptaddress]{revtex4}
\usepackage{graphicx}
\usepackage[latin1]{inputenc}
\usepackage{times}
\def\Ham{\mathcal{H}}
\newcommand{\ket}[1]{\left\vert{#1}\right\rangle}

\newcommand{\md}[1]{\vert{#1}\vert}

\newcommand{\moy}[1]{\langle{#1}\rangle}
\newcommand{\inter}[2]{\left\langle{#1}\vert{#2}\right\rangle}

\def\rwth{Institute for Theoretical Physics C, RWTH Aachen University, D-52056 Aachen, Germany}
\def\polytechnique{Centre de Physique Th\'{e}orique, Ecole Polytechnique, 91128 Palaiseau Cedex, France}
\def\geneve{DPMC-MaNEP, University of Geneva, 24 Quai Ernest-Ansermet, CH-1211 Geneva, Switzerland}
\def\queens{School of Physical Sciences, The University of Queensland, Brisbane, QLD, 4072, Australia}

\begin{document}

\author{G. Roux}
\email{roux@physik.rwth-aachen.de}
\affiliation{\rwth}
\author{T. Barthel}
\affiliation{\rwth}
\author{I. P. McCulloch}
\affiliation{\queens}
\author{C. Kollath}
\affiliation{\polytechnique}
\author{U. Schollw\"{o}ck}
\affiliation{\rwth}
\author{T. Giamarchi}
\affiliation{\geneve}

\date{\today}

\title{The quasi-periodic Bose-Hubbard model and localization in
  one-dimensional cold atomic gases}

\pacs{03.75.Lm, 61.44.Fw, 67.85.Hj, 71.23.-k}

\begin{abstract}
  We compute the phase diagram of the one-dimensional Bose-Hubbard
  model with a quasi-periodic potential by means of the density-matrix
  renormalization group technique. This model describes the physics of
  cold atoms loaded in an optical lattice in the presence of a
  superlattice potential whose wave length is incommensurate with the
  main lattice wave length. After discussing the conditions under
  which the model can be realized experimentally, the study of the
  density vs. the chemical potential curves for a non-trapped system
  unveils the existence of gapped phases at incommensurate densities
  interpreted as incommensurate charge-density wave phases.
  Furthermore, a localization transition is known to occur above a
  critical value of the potential depth $V_2$ in the case of free and
  hard-core bosons. We extend these results to soft-core bosons for
  which the phase diagrams at fixed densities display new features
  compared with the phase diagrams known for random box distribution
  disorder. In particular, a direct transition from the superfluid
  phase to the Mott insulating phase is found at finite $V_2$.
  Evidence for reentrances of the superfluid phase upon increasing
  interactions is presented. We finally comment on different ways to
  probe the emergent quantum phases and most importantly, the existence
  of a critical value for the localization transition. The later
  feature can be investigated by looking at the expansion of the cloud
  after releasing the trap.
\end{abstract}

\maketitle 

Disordered media are known to allow for the localization of waves in
many physical systems, both quantum and classical. As demonstrated by
Anderson~\cite{Anderson1958,Lee1985}, increasing disorder induces a
transition to an insulating state. The occurrence of this Anderson
transition strongly depends on the dimensionality of the system: in
one-dimension, a localized phase is expected as soon as disorder is
present~\cite{Abrahams1979}. One of the key question in the field of
strongly correlated systems is the interplay between interactions and
disorder. Using field theoretical methods~\cite{Giamarchi1987,
  Giamarchi1988}, it was shown that, for one dimensional systems of
bosons and fermions, interactions can lead to a
localization-delocalization transition. For one
dimensional~\cite{Giamarchi1987, Giamarchi1988} or higher
dimensional~\cite{Fisher1989} bosons, the combination of interactions
and disorder leads to a transition between a superfluid phase for
weakly repulsive bosons and a localized phase (Bose glass) for strong
repulsion. When an additional commensurate potential is present, there
is a competition between the three possible phases, namely the
superfluid (SF) phase, the Mott insulating (MI) phase, which occurs
for commensurate fillings and large interactions, and the so-called
Bose-glass (BG) phase, which is induced by disorder. Numerical
studies~\cite{Batrouni1990, Scalettar1991} supported the general
picture and provided phase diagrams~\cite{Prokofev1998, Rapsch1999} in
one dimension where mean-field theory fails. However, experimental
set-ups in solid state physics lack a good control of the interactions
and the disorder strength. More recently, cold atomic gases offered
the possibility of a fine-tuning of the hamiltonian parameters in
particularly clean set-ups. As a paradigm for strongly interacting
gases, the SF-MI phase transition was demonstrated using an optical
lattice~\cite{Bloch2007}. A fine-tuning of the disorder strength is
likewise conceivable. In this direction, several proposals were put
forward: the use of a laser speckle~\cite{Lye2005, Clement2005,
  Fort2005, Schulte2005, Clement2006, Clement2008, Chen2007}, the use
of heavy atoms, which provide a quasi-static potential for lighter
atoms~\cite{Paredes2005,Gavish2005}, and finally the addition of a
superlattice potential with a wave length incommensurate with that of
the lattice potential~\cite{Diener2001, Damski2003, Lye2007,
  Fallani2007}.

\begin{figure*}[t]
\includegraphics[width=0.95\textwidth,clip]{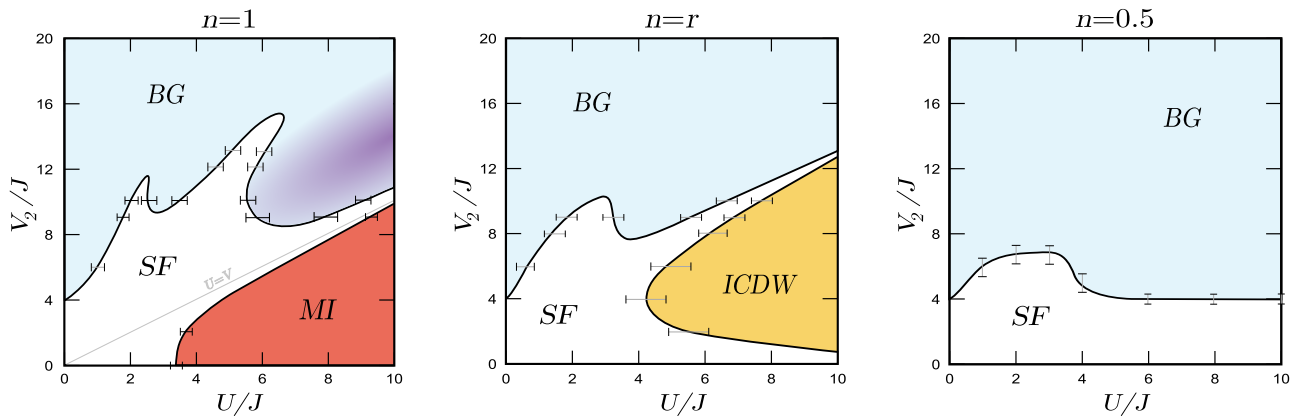}
\caption{(Color online) \emph{Phase diagrams of the bichromatic
    Bose-Hubbard model for densities $n=1$, $r$ (the ratio of the
    potential wave lengths) and $n=0.5$}. The diagrams are shown as a
  function of the interaction strength $U$ and the bichromatic
  potential strength $V_2$, both normalized by the hopping $J$ (lines
  are guides to the eyes). \emph{SF} stands for the superfluid phase,
  \emph{MI} for the Mott-insulating phase, \emph{BG} for the
  ``Bose-glass'' phase (meaning localized but with zero one-particle
  gap) and \emph{ICDW} for incommensurate charge-density wave phase.
  The $U=V_2$ line on the phase diagram with $n=1$ indicates the $J=0$
  limit for which the gap of the one-particle excitation vanishes.
  Black error bars are deduced from calculations averaging over the
  phase-shift $\phi$ (cf.  Sec.~\ref{subsec:energyscales}) and
  finite-size scaling (see Figs.~\ref{fig:XiCutAll} and
  \ref{fig:Reentrance} for details on the $n=1$ phase diagram). Grey
  error bars are roughly evaluated from calculations on systems with
  $L=35$ and fixed $\phi = 0$ (see Figs.~\ref{fig:MapN1} and
  \ref{fig:MapNR} for details). In the phase diagram with density
  $n=1$, the darker (violet) region in the BG phase is localized but
  could have a small gap which cannot be resolved numerically.}
\label{fig:phasediagrams}
\end{figure*}

This paper is devoted to the study of the latter situation, the
so-called bichromatic set-up, for which experiments have recently been
carried out~\cite{Lye2007, Fallani2007, Guarrera2007, Guarrera2008}.
The one-particle Schr\"{o}dinger equation with an incommensurate
lattice has been widely studied~\cite{Aubry1980, Simon1982,
  Thouless1983, Kohmoto1983, Kohmoto1983a, Diener2001} and was found
to exhibit anomalous diffusion properties~\cite{Zhong1995}. The main
result of these studies, as we will recall later on, is the existence
of a critical value of the potential above which localization occurs.
For the many-body physics, a weak-coupling treatment of the potential
was carried out using bosonization~\cite{Vidal1999}. Quasi-periodic
potentials were found to have an intermediate behavior between
commensurate ones, and disordered ones.  Exact numerical results on
the Bose-Hubbard model with a quasi-periodic potential already
exist~\cite{Roth2003, Bar-Gill2006, Louis2006} but are limited to
small systems and thus could not investigate the nature of the
transition nor their precise location.  The physics of the
Bose-Hubbard model with a \emph{periodic} superlattice has been
investigated~\cite{Buonsante2004, Buonsante2005, Rousseau2006} and a
``weakly superfluid'' phase at large potential depth was
found~\cite{Rousseau2006}. Very recently, Roscilde~\cite{Roscilde2007}
carried out a more detailed study using quantum Monte Carlo
calculations and a ``random atomic limit'' approach. This study gave
results on the bulk system for a special choice of parameters, and an
accurate description of the physics of the trapped cloud, focusing on
static observables. A particularly important point in which we go
beyond Ref.~\onlinecite{Roscilde2007} is the description of the phase
diagrams of the bulk system at fixed densities which are essential to
understand the interplay between the competing orders at stake.

Our main motivation is to address the shape of these fixed-density
phase diagrams for a one-dimensional system using the density-matrix
renormalization group (DMRG) algorithm (see section~\ref{sec:numerics}
for details), which results are interpreted within the framework of
the Luttinger liquid theory. We focus on the differences and
similarities between the deterministic bichromatic lattice potential
and a truly random one, usually consisting of a random box
distribution (RBD) and for which the phase diagrams without a trap are
known~\cite{Prokofev1998, Rapsch1999}. An account of our main results
are depicted in Fig.~\ref{fig:phasediagrams}, which gathers the phase
diagrams for three typical densities as a function of the interaction
strength $U$ and the disorder potential strength $V_2$ (see
Sec.~\ref{subsec:energyscales} and Eq.~(\ref{eq:hamiltonian}) for a
precise definition of the hamiltonian). $n$ is the density of bosons
and $r$ is the ratio of the employed lattice wave lengths, which
characterizes the incommensurability of the potential. A first
interesting result is that a finite $V_2 \geq 4$ is always required to
stabilize the BG phase. We must precise that the term BG is used to
call a localized phase which is compressible (with a zero one-particle
gap), but the detailed features of the BG phase of the bichromatic
potential differ from the usual RBD BG phase as will be discussed in
what follows.  Contrary to the RBD phase diagram, we argue, based on
numerical evidence, that there is no intervening BG phase between the
SF and the MI phase at density $n=1$. An incommensurate charge-density
wave (ICDW) phase -- referred to as the incommensurate band insulator
(IBI) phase in Ref.~\onlinecite{Roscilde2007} -- emerges at finite
$V_2$ for a density $n \simeq r$. Lastly, we observe that the larger
the density, the larger the extension of the SF phase is.

The paper is organized as follows: in section~\ref{sec:model}, we
first give the conditions under which the hamiltonian describing the
many-body physics simplifies into a simple lattice hamiltonian used
for numerical calculations. We then discuss one of the strongest
differences compared to a random box distribution which is the
emergence of plateaus in the density-chemical potential curve
(section~\ref{sec:plateaus}). We next discuss, in
section~\ref{sec:diagrams}, the competition between the disorder
potential and the interactions by computing the phase diagrams at
integer density one and for a density for which an ICDW plateau
occurs. Lastly, section~\ref{sec:dynamics} is dedicated to the
possible relevant experimental probes of localization by focusing on
the out-of-equilibrium dynamics of the system.

\section{The Bose-Hubbard model with an incommensurate superlattice}
\label{sec:model}

\subsection{Energy scales hierarchy: Validity of the model}
\label{subsec:energyscales}

\begin{figure}[t]
\includegraphics[width=0.75\columnwidth,clip]{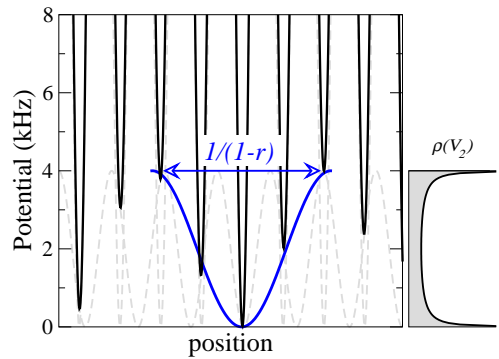}
\caption{(Color online) The bichromatic potential (full line) with the same
  parameters as in the experiment of Ref.~\onlinecite{Lye2007}. Dashed
  lines show the two beating standing waves from which the bichromatic
  potential originates. We observe that not only the depths of the
  potential wells fluctuate but so do their positions and their width.
  The energetic landscape displays wells of typical width $1/(1-r)$.
  The plot on the right-hand side shows the bounded distribution which
  behaves as $1/\sqrt{x(V_2-x)}$ and is thus peaked around 0 and
  $V_2$.}
\label{fig:typical-potential}
\end{figure}

This section gives qualitative arguments on the hierarchy of energy
scales which leads to a simple lattice hamiltonian that captures the
physics of cold bosonic atoms experiencing two optical lattice
potentials with wave vectors $k_1,k_2$ and amplitudes $V_1,V_2$.
Similar considerations were given recently in
Ref.~\onlinecite{Guarrera2007}. For the sake of clarity, we keep the
following discussion on the regime of experimental parameters under
which the lattice hamiltonian under study is valid. The potential
energy in a one-dimensional setup of two standing-waves is the Harper
potential
\begin{equation}
\label{eq:potential}
V(x) = V_1 \cos^2(k_1 x) + V_2 \cos^2(k_2 x + \phi)\;,
\end{equation}
which is sketched in Fig.~\ref{fig:typical-potential}. A constant
phase $\phi$ is introduced to shift the second lattice with respect to
the other, and the wave vectors $k_1$ and $k_2$ can take any value. We
work in the limit of a large depth $V_1 \gg E_{r1}$ for which we can
restrict ourselves to the lowest Bloch band ($E_{r1} = (\hbar
k_1)^2/2M$ is the recoil energy associated with the first lattice and
$M$ is the mass of the atoms) and in a situation where one intensity
is much larger than the other, $V_1 \gg V_2$. An exact derivation of
the lattice parameters of the hamiltonian should resort to numerical
calculations as described in Refs.~\onlinecite{Damski2003,
  Roscilde2007}. Our motivation is to evaluate the physical effect of
the perturbing potential to deduce the relative magnitudes of the
different terms.

To proceed, we neglect the effect of the trap on the local chemical
potential and displacements, meaning that we consider the realistic
situation for the bulk physics with $\lambda_1 \lambda_2 /
\vert\lambda_1 - \lambda_2\vert, \lambda_1, \lambda_2 \ll
\sqrt{\hbar/M\omega}$ with $\omega$ the trap frequency. If $V_2 = 0$, the
effective model is the Bose-Hubbard model~\cite{Bloch2007} with
hopping $J_0$ and on-site interaction $U_0$
\begin{equation}
\label{eq:bose-hubbard}
\Ham_0 = -J_0 \sum_j [b^{\dag}_{j+1} b_j + b^{\dag}_{j} b_{j+1} ] 
         +U_0 \sum_j n_j(n_j-1)/2\,.
\end{equation}
$b^{\dag}_j$ is the operator that creates a boson at site $j$
corresponding to the minimum of the lattice potential. The local
particle number operator reads $n_j = b^{\dag}_{j} b_{j}$. The
dependence of the parameters $J_0$ and $U_0$ upon $V_1, E_{r1}, k_1$
and the scattering length $a$ can be evaluated numerically or
analytically in this limit~\cite{Bloch2007}. We now qualitatively
discuss the effect of $V_2(x)$ to the lowest order in $\varepsilon =
V_2/V_1$.

\emph{Perturbation of the chemical potential} -- First, to zeroth
order in $\varepsilon$, the minima of $V(x)$ are located at $k_1 x_j =
\pi j + \pi/2$ with $j$ integer and $\pi/2$ can be absorbed in the
redefinition of $\phi$. Since $V_1(x_j) = 0 + O(\varepsilon^2)$, we
have:
\begin{equation*}
V(x_j) = V_2 \cos^2 \left( r\pi j + \phi\right) 
= \varepsilon \frac {V_1} 2 [1 + \cos \left( 2r\pi j + 2\phi\right)]
\end{equation*}
The important number which characterizes this bichromatic potential is
the ratio of the wave vectors $r = k_2 / k_1$. If $r$ is a rational
number $p/q$, the hamiltonian is $q$-periodic. For $r$ irrational, it
has no translational invariance and the $V(x_j)$ can take any value in
$[0,V_2]$ in a deterministic way: the resulting bounded distribution
is sketched in Fig.~\ref{fig:typical-potential}. The chemical
potential thus shares features with a one-dimensional quasi-crystal.
The order of magnitude of the coefficient of this term is of course
$V_2 = \varepsilon V_1$. This term can be larger than $J_0$ or $U_0$,
even though $\varepsilon \ll 1$, because of the factor $V_1$. Taking
the parameters of Ref.~\onlinecite{Lye2007}, one finds that
$\varepsilon \simeq 0.003$--0.12 while $V_2/J_0 \simeq 2.6$--53.3.

\emph{Perturbation of the hopping} -- It is difficult to treat the
perturbation of the hopping exactly because one needs to know the
non-trivial shape of the perturbed Wannier functions. However, we
expect the hopping to be perturbed mainly because of the displacement
of the local minima and because tunneling depends exponentially on the
distance. We only consider the term associated with the perturbation
of the minima $x_{j}$ and $x_{j+1}$ and assume a typical exponential
dependence~\cite{Bloch2007} for the hopping $J_{j,j+1} \sim
e^{-h(x_{j+1}-x_j)}$ with $ h=\frac {k_1}{2} \sqrt{\frac{V_1}{E_r}}$,
valid for $V_1 \gg E_{r1}$. The modulation induces a slight
fluctuation $\delta x_j$ at site $j$ which, to the lowest order in
$\varepsilon$, reads:
\begin{equation*}
\delta x_j = -\frac{1}{2k_1} \varepsilon r\sin\left(2r\pi j +2\phi\right)\;,
\end{equation*}
so that the distance between two neighboring sites is altered as:
\begin{equation*}
x_{j+1} - x_j = \frac {\pi} {k_1}
- \varepsilon \frac{r}{k_1} \sin(\pi r)\cos\left(r\pi (2j+1) +2\phi\right)\;.
\end{equation*}
Hence, to the lowest order in $\varepsilon$, we may write for this term:
\begin{equation*}
J_{j,j+1} = J_0\left[1 + \varepsilon A \cos\left(r\pi (2j+1) +2\phi\right)\right]
\end{equation*}
with, up to approximations, $A = \sqrt{\frac{V_1}{E_{r1}}} \frac r 2
\sin(\pi r)$.  We write $J_2 = \varepsilon A J_0$. Even though $A$
could be large because of $\sqrt{V_1/E_{r1}}$, the factor
$\varepsilon$ ensures that $J_2$ can be made much smaller than $J_0$.
More precisely, taking the parameters of Ref.~\onlinecite{Lye2007} and
using the above approximation, one finds $J_2/J_0 \simeq 0.002$--0.1,
the latter occurring for very large $V_2$, much larger than the ones
used in this paper. Numerical calculations of the hamiltonian
parameters~\cite{Damski2003, Roscilde2007} confirm that the magnitude
of $J_2$ is small within our approximations. Hence, we will shorten
the notation to $J \equiv J_0$ from now on. Another feature which
results from this approximation is that the $J_{j,j+1}$ have the same
typical fluctuations $\cos(2\pi rj)$ as the $V(x_j)$ which is observed
numerically in Ref.~\onlinecite{Roscilde2007}, yet for rather large
$\varepsilon$.

\emph{Perturbation of the local interaction} -- In the deep well
limit, the bare interaction $U_0$ is obtained by the
relation~\cite{Bloch2007}:
\begin{equation*}
U_0 = \sqrt{\frac{8}{\pi}} k_1 a E_{r1} \left(\frac {V_1}{E_{r1}}\right)^{3/4}
\end{equation*}
where $a$ is the scattering length. Note that the ratio $U_0/J_0 \sim
\exp(2\sqrt{V_1/E_{r1}})$ increases exponentially with the lattice
depth~\cite{Bloch2007}. This result can be obtained by approximating
the bottom of the cosines with a parabola and using gaussian Wannier
functions as the simplest approximation. The fact that $U_0$ increases
with $V_1$ simply corresponds to the squeezing of the parabola. This
squeezing may also be changed at first order by $V_2$. To give a rough
estimate, we can compute the second derivative of
Eq.~(\ref{eq:potential}) and obtain for the on-site interaction:
\begin{equation*}
U_0 + U_2 \cos(2r\pi j + 2\phi) \text{ with } U_2 
   = \frac 3 4 \varepsilon r^2  U_0\;. 
\end{equation*}
Here again, since $\varepsilon$ can be tuned to be very small, one can
work within the $U_2 \ll U_0$ regime. The perturbation of the on-site
interaction can thus be neglected and we will use the shorter notation
$U \equiv U_0$ in the following. We also note that the fluctuations
of the local interactions have roughly the same cosine dependence as
the chemical potential.

To conclude, in the deep well limit $V_1 \gg E_{r1}$, the
following hierarchy of energy scales
\begin{equation*}
J_2, U_2 \ll U_0,J_0,V_2 \ll V_1
\end{equation*}
can be realized experimentally. Thus, we assume the corresponding
lattice model for the bichromatic optical lattice:
\begin{eqnarray}
\nonumber
\Ham &=& -J\sum_j [b^{\dag}_{j+1} b_j + \text{h.c.} ] + U \sum_j n_j(n_j-1)/2 \\
\nonumber
&& + \frac {V_2} 2 \sum_j [1 + \cos \left( 2r\pi j + 2\phi\right)] n_j \\
\label{eq:hamiltonian}
&& + \frac {\omega^2} 2 \sum_j (j-j_0)^2 n_j
\end{eqnarray}
with $j_0 = (L+1)/2$ the center of the trap. In what follows, results
for the phase diagrams are computed with $\omega=0$.  The trap
confinement is added in a few illustrating figures and more
importantly, for the preparation of the out-of-equilibrium state in
the study of dynamics (Sec.~\ref{sec:dynamics}).

We can briefly comment on the distribution of the on-site potential
energies as it is the first difference with the RBD. We shall use the
short-hand notation for the bichromatic potential $V_2(x_j) \equiv V_j
= V_2 [1 + \cos( 2r\pi j + 2\phi)]/2$. The distribution of the $V_j$
with an irrational $r$ behaves as $1/\sqrt{x(V_2-x)}$ which diverges
close to 0 and $V_2$ and is symmetrical with respect to $V_2/2$ (see
Fig.~\ref{fig:typical-potential}) but is relatively flat at the
center. Thus, this distribution qualitatively lies in between a RBD
and a binary one. Its auto-correlation function reads:
\begin{equation*}
\overline{V_j V_{j+d}} - \overline{V_j}^2 = (V_2)^2 \cos(2 \pi r d + 2 \phi)/8
\end{equation*}
where the over-bar means averaging over all sites $j$. The potential
is thus deterministic and correlated. Though trivial, this remark
stresses the fact that the very features of the localization mechanism
under study originates from the quasi-periodicity rather than the
distribution itself. For instance, an uncorrelated disordered
potential with the same distribution would induce localization as soon
$V_2$ is finite, which is not true for the bichromatic one. As
sketched in Fig.~\ref{fig:typical-potential} by black and dashed grey
lines, wells develop over a characteristic length scale $1/(1-r)
\simeq 4.4$ sites which comes from the beating of the two periods 1
and $r$ of the two lasers.

Working with finite systems raises the question of taking the
thermodynamical limit. The system length $L$ is given in units of the
first lattice spacing $\lambda_1/2$. First, to what extent can an
irrational number $r$ be approximated by a rational number?  This can
be answered by looking at its continuous fraction
decomposition~\cite{wikicontinuedfraction} which gives the successive
best rational approximations. From Ref.~\onlinecite{Lye2007}, $r =
830.7/1076.8 = 0.77145245\ldots$ is a realistic ``irrational''
parameter as 8307 and 10768 are coprimes. The successive best rational
approximations are $3/4,7/9,10/13,17/22,27/35,908/1177,\ldots$ which
gives the lengths $L=13,22,35,1177$, best ``fitting'' the potential
for non-trapped systems. As $27/35$ is already a fairly good
approximation of the ``irrational'' $r$ of the experiments, multiples
of 35 such as 70, 105, can be used as well. We will also use other
lengths $L$ and we have checked that the physics does not change
qualitatively if the system size does not perfectly fit the potential.
The fact that 35 is a rather large period ensures that $r$ is not too
close to a simple fraction which would induce strong commensurability
effects on finite systems. In what follows, we choose to work with the
experimental parameter $r = 0.77145245$ as Roscilde
did~\cite{Roscilde2007} to be as close as possible to the experiments
but we expect the general picture to remain true for any irrational
number. Furthermore a phase shift $\phi$ enters in the hamiltonian
and, though we expect the potential to be self-averaging for fixed
$\phi$, averaging over $\phi$ can help recover the thermodynamical
limit. This averaging will be denoted by $\moy{}_{\phi}$ in the
following. As experimental set-ups generally consist in an assembly of
one-dimensional cigar-shaped clouds with different lengths (see Fig.~3
of Ref.~\onlinecite{Fallani2007} for instance), clouds with different
lengths would effectively experience a different phase shift $\phi$.
Furthermore, from one shot to another, the tubes experience slightly
different potentials. This is due to the difficulty to lock the
position of the cloud in the trapping and optical lattice potentials
over several shots. Consequently, $\phi$ may fluctuate from one
preparation to another.

The last crucial parameter in the physics of the system is the density
which plays an important role as we will see. For a non-trapped
system, we use the notation $n = N / L$ with $N$ the total number of
bosons which is kept fixed as we work in the canonical ensemble.
For a trapped system, the local density varies as one moves away from
the middle of the trap and the thermodynamical limit is recovered for
$\omega \rightarrow 0$ keeping $N\sqrt{\omega}$ constant.
Roscilde~\cite{Roscilde2007} gave a detailed analysis of the static
properties in the presence of a trap. Our focus is more on the phase
diagram of the model which is always understood to be in the
thermodynamical limit. More details with respect to experimental
probing will be given in section~\ref{sec:dynamics}. All results of the
paper are for zero temperature.

\subsection{Low-energy approach: bosonization}
\label{sec:bosonization}

We briefly review known results from the low-energy approach
(close to a hydrodynamic description) which will be useful for the
interpretation of the numerics and offer a complementary point of view
on the physics. The low-energy physics of interacting bosons in 1D are
described using Haldane's harmonic fluid approach~\cite{Haldane1981,
 Cazalilla2004, Giamarchi2004} in which the density operator is
expanded as:
\begin{equation}
\rho(x) = \left[n - \frac 1 {\pi} \nabla \phi(x) \right] 
         \sum_{p=0,\pm 1, \pm 2,\ldots} e^{ip(2\pi n x- 2\phi(x))}
\end{equation}
where $n$ is the boson density which encompasses the lattice spacing
$d$. The effective hamiltonian of the system has generically a
quadratic part which includes a kinetic energy term $\sim \Pi^2$, with
$\Pi=\frac{1}{\pi}\nabla\theta$ the conjugate field of $\phi$ (the
commutation relations $[\phi(x),\Pi(x')]=i\delta(x-x')$ and
$[\theta(x),\frac{1}{\pi}\nabla \phi(x')]=i\delta(x-x')$ hold), and a
density-density like interaction term $\sim (\nabla \phi)^2$.  Two
Luttinger parameters $u$ and $K$ give a simple parameterization of the
quadratic part of the hamiltonian:
\begin{equation}
\label{eq:Bos-Hamiltonian}
\Ham = \int \frac{dx}{2\pi} \left\{ uK (\pi \Pi)^2 
                         + \frac u K (\nabla \phi)^2 \right\}
\end{equation}
where $u$ has the dimension of a velocity and $K$ is dimensionless.
For free bosons, only the first term remains which would formally
correspond to the $K \rightarrow \infty$ limit and $u = \sqrt{n \pi/
M}$ is the sound velocity. Taking into account a local interaction
$\frac U 2 \rho(x)^2$ like in the Bose-Hubbard model, the Luttinger
parameters read $u=\sqrt{n U / M}$ and $K = \pi \sqrt{n / M U}$ in the
limit $U \ll J$. When interactions are large, higher harmonics in the
density operator have to be taken into account to describe correctly
the local fluctuations and not only the long-distance ones. In the $U
= \infty$ limit, i.e.~for hard-core bosons (HCB), one obtains $K=1$ as
one would find for free fermions. The strong interaction, i.e.~the
second term in Eq.~(\ref{eq:Bos-Hamiltonian}), acts as a Pauli
exclusion term. We thus generically have $1 \leq K < \infty$ for
on-site repulsive interactions. The effective hamiltonian
(\ref{eq:Bos-Hamiltonian}) provides the general low-energy description
of the SF phase which can undergo various instabilities.

By bosonizing the standard Bose-Hubbard model~(\ref{eq:bose-hubbard}),
commensurability effects can arise from the higher
harmonics~\cite{Giamarchi2004}:
\begin{equation*}
\rho(x)^2 = n^2 + \frac {1}{\pi^2} (\nabla \phi) ^2 
+ n^2 \sum_{p > 1} \cos[2\pi p n x - 2p\phi(x)] + \ldots
\end{equation*}
>From studying the renormalization group (RG) flow equations, it is
known that cosine terms such as
\begin{equation}
\label{eq:interactionpot}
U \int dx \cos[2\pi p n x - 2p\phi(x)]
\end{equation}
can lock the density field $\phi$ and induce a
commensurate-incommensurate transition~\cite{Dzhaparidze1978} (C-IC)
depending on the density and of $K$. Working at fixed density and
varying interactions, such a term is relevant only if the density
satisfies the commensurability condition $p n = 1,2,\ldots$ and if $K
< K_c$ with $K_c = 2/p^2$. The opening of the gap follows a
Kosterlitz-Thouless~\cite{Kosterlitz1973} (KT) law $\Delta_c \sim
\exp(-A/\sqrt{U-U_c})$ with $A$ a constant and $U_c$ the critical
value. Working at fixed interaction and varying the density, the
commensurate phase is obtained for $K_c = 1/p^2$. For instance, for $p
= 1$, integer densities $n = 1,2,3,\ldots$ allow for a Mott phase for
$K$ below $K_c = 2$. For $p=2$, charge-density wave phases can appear
for half-integer densities but nearest-neighbor repulsion are
required~\cite{Giamarchi2004, Giamarchi1997, Kuhner2000} to get $K <
1$. It is important to note that such cosine effective potential terms
\emph{effectively} arise from the interactions. The transition towards
the charge density-wave (CDW) state with one atom every two
sites~\cite{Kuhner2000} is associated with a spontaneous breaking of
the translational symmetry.  The other possibility to generate Mott
phases is to \emph{artificially} introduce a cosine chemical potential
which directly couples to the density. Similarly, a CDW phase induced
by an external potential is associated with an explicit breaking of
the translational symmetry. This latter solution is possible in cold
atoms by adding a superlattice.

\emph{Effect of a superlattice potential} -- We first
consider a cosine potential $V_2(x) = V_2 \cos(Q x)$ which has only one
Fourier component at wave vector $Q = 2\pi r$ with $r$ rational. The
additional term reads
\begin{equation}
\label{eq:bichropot}
\int dx V_2(x) \rho(x) 
= \frac{V_2}{2} \int dx \cos[(2\pi n \pm Q)x - 2\phi(x)] + \ldots
\end{equation}
As seen previously, such terms may induce a C-IC transition when
increasing $V_2$ if the condition $n \pm r \in \mathbb{Z}$ is
satisfied. In particular, the superlattice potential can become
relevant for the densities $n = r, 1-r, 1+r, 2-r,\ldots$. Higher
harmonics can be generated, as discussed in
Sec.~\ref{sec:hcb-plateaus}. If the potential term is not relevant,
the Luttinger parameter $K$ is, however, renormalized to a lower value
by the potential as it happens with interactions. Such commensurate
potentials have for instance been studied in the context of Mott
transitions~\cite{Giamarchi1997} and of magnetization
plateaus~\cite{Arlego2001}. The physics of cold atoms with induced
commensurate CDW phases was studied in detail in
Ref.~\onlinecite{Rousseau2006}. Vidal~\textit{et al.}~\cite{Vidal1999}
generalized this result to irrational $r$. For the case of a
quasi-periodic potential, the critical value $K_c$ remains equal to 2
if the density approximately satisfies the relation $n \pm r \in
\mathbb{Z}$. If the density does not fulfil this condition but remains
close enough, an insulating phase can be reached but for smaller
critical value (for a spin-less fermion model, $K_c \simeq 1$ was
found from RG).

\emph{Disorder with a random box distribution} -- From
Refs.~\cite{Giamarchi1987, Giamarchi1988, Fisher1989}, the main result
is that the potential is relevant below the critical value $K_c =
3/2$, whatever the density. The resulting Bose-glass phase has no
one-particle gap but an exponentially decaying one-particle Green's function due
to localization. The correlation length scales according to $\xi \sim
\exp(-A / (V_2 -V_2^c))$ where $V_2^c$ is the critical value for the
transition.

\subsection{Numerical methods}
\label{sec:numerics}

The hard-core bosons physics can be solved exactly using a
Jordan-Wigner transformation which maps the model onto free fermions
with boundary conditions that depend on the number of bosons. As the
method has been widely described in the literature, we refer the
reader to Refs.~\cite{Rigol2005, Roscilde2007}. This method is also
used to investigate the out-of-equilibrium properties~\cite{Rigol2005}
of the cloud in section~\ref{sec:dynamics}.

We use the DMRG algorithm~\cite{White1992, White1993, Schollwoeck2005}
to treat the soft-core Bose-Hubbard model~(\ref{eq:hamiltonian}). For
disordered systems, sweeping has proven to be particularly important
to get reliable results~\cite{Schmitteckert1998, Rapsch1999}. DMRG has
also been used to investigate the physics of quasi-periodic electronic
systems~\cite{Hida2001, Schuster2002}. Our implementation is based on
a matrix-product state variational formulation~\cite{McCulloch2007}
which enables us to start sweeping from any state. In practice, we
have started from either a random or a classical state (where the
particles are located according to the $J=0$ limit of the hamiltonian)
contrary to the usual warm-up method. The algorithm works in the
canonical ensemble (fixed number of particles $N$) and at zero
temperature. We typically use from 200 to 400 kept states. The number
of bosons allowed on-site is usually fixed to $N^{\text{bos}}=4$ but
results for densities larger than one have also been checked with up
to $N^{\text{bos}}=6$.  For $U \geq 1$ and $V_2 \leq 20$, the
classical distribution of particles does not have more than 4 bosons
per sites.

A drawback of this variational method is the occasional tendency to
get trapped in an excited (metastable) state with a slightly higher
energy that is difficult to distinguish numerically from the
ground-state. Indeed, the usual measures of the convergence of DMRG,
the discarded weight and the variance $\moy{(\Ham-E)^2}$ are very
small for these states. Systematic tests have been carried out in the
$U \rightarrow \infty$ limit. Starting from the classical state
improves convergence for small densities or close to one at large
$V_2$ as one would expect intuitively. Below $V_2 \simeq 4$,
convergence is always good which can be related to the physics of the
systems as the potential does not induce localization in this regime.
In the case of soft-core bosons, we expect an enhancement of quantum
fluctuations at finite $U$ to help the particles redistribute more
easily. Such equilibration is rendered very difficult for HCB as for
strong $V_2$, local densities can be very close to one. Most of the
data have been obtained for $U \lesssim V_2$. Furthermore, relying on
the variational principle, we can use the smallest of the two energies
obtained from starting either from the classical or a random state.
Lastly, the coherence of the results obtained from observables
computed independently, such as the correlation length and the
one-particle gap (see section~\ref{sec:diagrams}), supports the good
convergence of the algorithm.

\section{Density plateaus: Mott and incommensurate charge density wave
 phases}
\label{sec:plateaus}

This section describes the relation between the density $n$ and the
chemical potential $\mu$ for a non-trapped cloud. The motivation is to
find first the location of the compressible and incompressible phases.
The chemical potential is computed via
\begin{equation*}
\mu(N) = E_0(N) - E_0(N-1)
\end{equation*}
where $E_0(N)$ is the ground-state energy with $N$ bosons. If a
plateau emerges in the $n(\mu)$ curve, its width is directly related
to the one-particle gap defined by
\begin{equation*}
\begin{split}
\Delta_c &= E_0(N+1) + E_0(N-1) -2  E_0(N)\\
& = \mu(N+1)-\mu(N)
\end{split}
\end{equation*}
Lastly, the compressibility of the system $\kappa = \partial n /
\partial \mu$ is evaluated through its discretized expression as 
\begin{equation}
\label{eq:compressibility}
\kappa^{-1} = L [ E_0(N+1) + E_0(N-1) -2  E_0(N) ] \;.
\end{equation}
For a Luttinger liquid, the compressibility is simply related to the
Luttinger parameter and the sound velocity:
\begin{equation}
\label{eq:Luttinger-compressibility}
\kappa = \frac{K}{\pi u}\;.
\end{equation}
The compressibility naturally vanishes in a plateau phase.

\subsection{Plateaus in the hard-core boson limit}
\label{sec:hcb-plateaus}

\begin{figure}[t]
\includegraphics[width=0.75\columnwidth,clip]{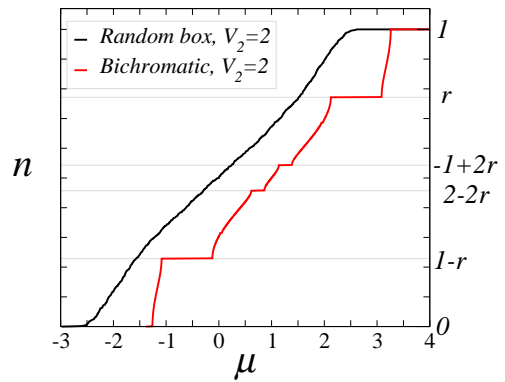}
\caption{(Color online) Comparison of $n(\mu)$ for a random and a
 bichromatic potential (irrational $r$) for hard-core bosons.
 Plateaus open for the bichromatic potential, the main ones being at
 $n=r$ and $1-r$.}
\label{fig:HCB-disorder-bichro}
\end{figure}

Following Ref.~\onlinecite{Fisher1989}, setting $J=0$ gives insight
into the $J \ll U$ physics. This gives the width $U-\max(V_j)+\min(V_j)$
of the various Mott plateaus centered at $\mu/U = 0.5, 1.5,
2.5,\ldots$. This is due to the fact that in the limit $J=0$ one can
reorder the energies by increasing values and therefore the $n(\mu)$
curve which is the integrated density of states is simply linear
between Mott plateaus for the random box distribution and $U >
\max(V_j)-\min(V_j)$. For a bichromatic lattice, we have $\mu = V_2
\sin(\pi n / 2)$. What happens when $J$ is small but finite? The
density of states evolves smoothly with $J$ for the random box
distribution (see Fig.~\ref{fig:HCB-disorder-bichro}).  For $V_2 = 0$,
the bandwidth which develops between the Mott plateaus has a width $4J$
and a cosine relation can be observed~\cite{Batrouni1990} because Mott
sub-bands with cosine dispersion are well separated. On the contrary,
for the bichromatic lattice intermediate plateaus appear as soon as
$J$ is non-zero. This behavior is reminiscent of the situation for
rational $r$ and was discussed extensively for free fermions, which in
our case would be equivalent to the HCB limit.

\begin{figure}[t]
\includegraphics[width=0.75\columnwidth,clip]{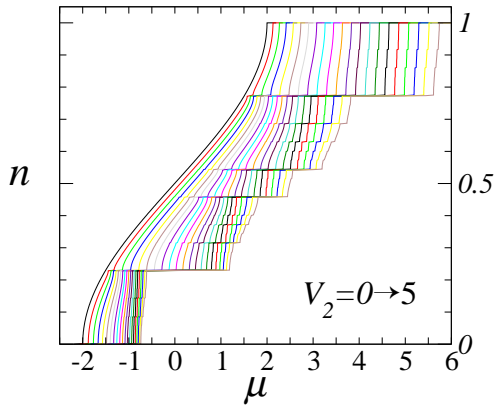}
\caption{(Color online) Opening of plateaus in the $n(\mu)$ curves
  with $V_2$ for hard-core bosons ($V_2$ is increased by steps of
  $J/4$ and is in units of $J$, as $\mu$). Above the critical value
  $V_2^c = 4$, the curves acquire a devil's staircase-like behavior:
  plateaus become dense.}
\label{fig:HCB-limit-irrational}
\end{figure}

The energy spectrum and the wave function properties have been widely
studied in the literature~\cite{Aubry1980, Simon1982, Kohmoto1983,
 Kohmoto1983a, Thouless1983, Barache1994}. It
was shown that gaps open in the energy spectrum. If $r$ is rational,
there is a finite number of gaps. If $r$ is irrational, there is an
infinite number of gaps at large $V_2$, the width of which strongly
depends on $p$ and $q$ if one writes $n = p/q$ and gets larger as
$V_2$ increases. We here recall the method usually followed: these
gaps are studied by $m$ successive approximations $r_m=p_m/q_m$ of the
irrational number $r$. For a given $m$, the potential is $q_m$
periodic and we can use Bloch's theorem on super-cells of length
$q_m$. The one-particle Schr\"{o}dinger equation of the
hamiltonian~(\ref{eq:hamiltonian}) reads:
\begin{equation}\label{eq:free-hamiltonian}
  -J( \psi_{j+1} + \psi_{j-1}) + [V_j - E] \psi_j = 0\,.
\end{equation}
Using $V_{j+q_m} = V_j$ and Bloch's theorem $\psi_{j+q_m} = e^{ikq_m}
\psi_j$, the spectrum is obtained by solving the determinant of size
$q_m$:
\begin{equation*} 
   \begin{vmatrix}
   V_1-E & -J       &        & -Je^{-ikq_m} \\
     -J     & V_2-E & -J     &        \\
            & -J       & \ddots &    -J     \\
   -Je^{ikq_m}&          &    -J   &  V_{q_m}-E \\
   \end{vmatrix}
   = 0
\end{equation*}
For $q_m = 2$, this is the simple band folding mechanism which opens a
gap at $n=1/2$, with a doubling of the unit cell. More generally, at
most $q_m -1$ gaps appear in the spectrum made of $q_m$ bands
$\mathcal{E}_{1,\ldots,q_m}(k)$ with $k \in [-\pi/q_m, \pi/q_m]$.
Examples of effective dispersion relations for the bichromatic
potential can be found in Sec.~\ref{sec:dynamics}.
Fig.~\ref{fig:HCB-limit-irrational} displays the opening of the
plateaus for HCB with $V_2$. A simple real-space interpretation can be
given for the main plateau at $n=1-r$: it amounts to fill each well of
size $1/(1-r)$ with one particle (see
Fig.~\ref{fig:typical-potential}). The plateau at $n=r$ is simply
obtained with the same argument with holes instead of particles.
Putting two particles (holes) in each well can lead to plateau at
densities $2(1-r)$ and $1-2(1-r)=-1+2r$. The fact that the main
plateaus at $n=r,1-r$ develop as soon as $V_2$ is turned on is
expected from the bosonization arguments of
Sec.~\ref{sec:bosonization}, since $K(V_2=0) =1 < K_c=2$ for HCB. From
a momentum space point of view, these opening are associated with
umklapp processes with a momentum transfer $Q$ which, modulo $2\pi$,
gives back the conditions $n=r,1-r$. In perturbation theory, processes
with larger momentum transfers can be obtained from
Eq.~(\ref{eq:bichropot}) with higher order terms in $V_2$. For
instance, to second order, terms with transfers $2Q$ (corresponding to
$n=2-2r$ and $-1+2r$) will appear if $K<1/2$. Consequently, a finite
$V_2$ is required to stabilize these plateaus (see also
Fig.~\ref{fig:K4HCB}). As $V_2$ is increased, such processes break the
spectrum up and make it point-like for the critical value $V_2=4$
which is beyond this weak-coupling bosonization interpretation.
Lastly, we note that this is particular to the Harper model. For the
Fibonacci chain~\cite{Vidal1999}, the Fourier transform of the
potential is already dense at small $V_2$. In our situation, the
Fourier spectrum gets denser as $V_2$ is increased.

\subsection{Plateaus for soft-core bosons}

\begin{figure}[t]
\includegraphics[height=0.65\columnwidth,clip]{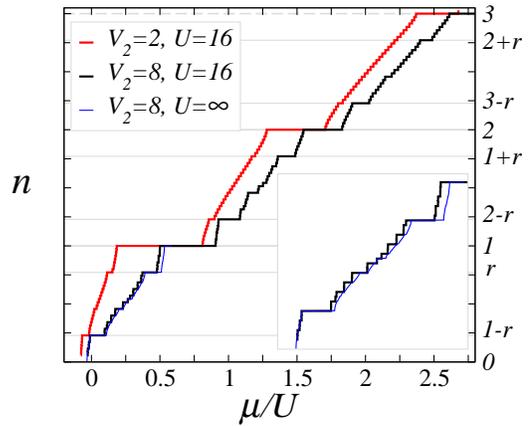}
\caption{(Color online) $n(\mu)$ for soft-core bosons with a large
  interaction parameter $U=16$ at small and large $V_2$ computed on a
  system with $L=35$ and a fixed phase-shift $\phi=0$. \emph{Inset:}
  curves at low density show the comparison between hard-core bosons
  ($U=\infty$) and soft-core ones, which is good up to finite size
  effects.}
\label{fig:N_of_mu_largeU}
\end{figure}

We now consider the case of a finite interaction $U$. First of all,
the hard-core boson limit is likely to give the correct qualitative
behavior for large $U$. Indeed, at low densities, an interaction $U$
slightly larger than $V_2$ might be sufficient to recover the HCB
physics as multiple occupancies are already strongly suppressed.
Densities larger than one are allowed for soft-core bosons. For large
$U$, we expect to find plateaus in between each Mott plateaus. One
may recover the hard-core bosons band folding mechanism inside each
Mott sub-bands (or at least for the lowest ones). These simple
observations are coherent with the large $U$ numerical data displayed
in Fig.~\ref{fig:N_of_mu_largeU}. A comparison with HCB results is
provided in the inset of Fig.~\ref{fig:N_of_mu_largeU} which proves
that $U=16$ is sufficiently large to reproduce the HCB physics within
the first three Mott sub-bands.

\begin{figure}[t]
\includegraphics[height=0.63\columnwidth,clip]{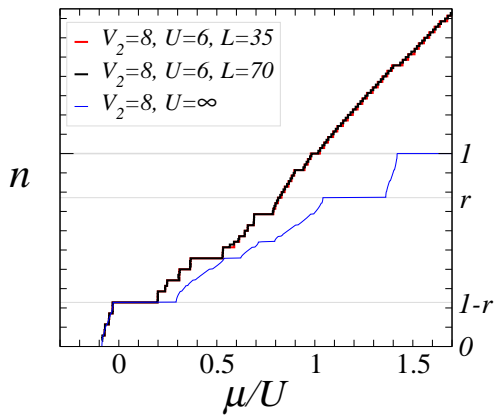}
\caption{(Color online) Plateaus with a large $V_2 \sim U$. Some of
  the largest plateaus found do not correspond to the hard-core boson
  limit. The two sizes $L=35$ and 70 give an idea of the (weak) finite
  size effects.}
\label{fig:N_of_mu_U6V8}
\end{figure}

Fig.~\ref{fig:N_of_mu_U6V8} gathers the results when $U \leq V_2$,
unveiling a more surprising behavior. As discussed previously, we
expect the HCB behavior to account for the low-density part of the
curve, which is actually observed through the rather large width of
the $n=1-r$ plateau. Indeed, because this plateau corresponds to one
particle in each well, the effect of interactions is restricted to
virtual processes. For higher densities, a large compressible phase is
obtained, manifested by the smooth increase of the density. From a
phenomenological point of view, adding atoms fills the well minima.
Since $U$ is not too large, the effective potential coming from the
combination of the interaction and the superlattice potential gets
smoother and smoother. Consequently, the associated gain in kinetic
energy favors a compressible and actually a superfluid state as we
will see in Sec.~\ref{sec:delocalization} where delocalization by
increasing the density is discussed. In between those two regimes, the
behavior is non-trivial. Strikingly, some plateaus existing in the HCB
limit totally disappear, such as the $n=r$ plateau, while others
acquire a larger width. Having gaps whose size increases when
interactions are reduced is something rather counter-intuitive. These
plateaus result from the interplay of the potential and the
interactions. A real-space picture was given by
Roscilde~\cite{Roscilde2007} following a random atomic limit:
considering wells of typical size $1/(1-r)$ separately, the fine
structure of the energy levels for each number of atoms inside the
wells depends strongly on $U,V_2$ and also on $J$. Connecting wells
with $J$ allows for the computation of the integrated density of
states which is $n(\mu)$. Though physically enlightening, this
approach is quantitatively correct for rather small densities. Since
the observed plateaus stem from the interplay of the interactions and
the potential, we call the corresponding plateau phase an
incommensurate charge density-wave phase. They appear to be the
extension of both the Mott and the incommensurate HCB phases at
smaller $U$. Bosonization explains, at least qualitatively, the
mechanism of plateaus opening in the HCB limit by considering high
order perturbative terms coming from Eq.~(\ref{eq:bichropot}), which
gives for instance the first two densities $r,1-r$ and $2(1-r),-1+2r$.
At finite $U$ and when $U \sim V_2$, the situation is more involved as
both terms should be treated non-perturbatively and on an equal
footing. Predicting the observed densities at which these ICDW phases
occur is thus beyond the perturbative approach.

\section{Localization induced by interactions or disorder: Phase
  diagrams}
\label{sec:diagrams}

We have seen that contrary to the standard random box situation, there
is not only one phase (either the BG or the SF) between the MI phases
but a succession of phase transitions as the chemical potential is
increased. This renders the usual~\cite{Fisher1989} interpretation of
the phase diagram in the $(\mu,J/U)$ plane for a fixed ratio $V_2/U$
rather strenuous~\cite{Roscilde2007} as the succession of phase
transitions breaks it up into many compressible and incompressible
pieces. Thus, we prefer to work at fixed density and varying the two
competing parameters $V_2/J$ and $U/J$. These phase diagrams were
first sketched numerically in Refs.~\cite{Roth2003, Louis2006,
  Bar-Gill2006} but on very small systems and without a discussion of
the boundaries and the nature of the transitions. We here provide a
more precise determination, in particular, by using scaling over
different sizes and averaging over $\phi$ when necessary. We now
describe more precisely the various observables used to sort out the
phases.

\subsection{Observables}
\label{sec:observables}

In addition to the compressibility, we need further observables to
sort out the different phases realized in the bichromatic set-up.  The
first natural one is the superfluid density $\rho_s$. It can be
computed using twisted boundary conditions:
\begin{equation}
\rho_s = 2\pi L \frac {E_0^{\text{apbc}} - E_0^{\text{pbc}}} {\pi^2} 
\end{equation}
where the ground-state energies are computed for periodic (pbc) and
anti-periodic (apbc) boundary conditions. With this definition,
$\rho_s$ actually matches the superfluid stiffness. Other
definitions~\cite{Rapsch1999} contain the density of particles $n$ as a
prefactor. The superfluid density is zero in the BG, ICDW and MI
phases and finite only for the SF phases. In a Luttinger liquid, the
superfluid density is directly related to the Luttinger parameters
through
\begin{equation}
\rho_s = u K\;.
\end{equation}
Combined with Eq.~(\ref{eq:Luttinger-compressibility}), $K$ can then
be computed using $K = \sqrt{ \pi \rho_s \kappa}$. This numerical
evaluation only requires the calculation of energies. $K$ can be
independently extracted from correlation functions. For instance, the
one-particle density-matrix or bosonic Green's function reads
$\moy{b^{\dag}_i b_j}$ where $\moy{}$ indicates the expectation value
in the ground-state. Following Ref.~\onlinecite{Kollath2004}, we
extract the contribution of the phase $\theta(x)$ fluctuations by
dividing it by the local inhomogeneous densities $n_i$:
\begin{equation}
\label{eq:green-function}
G(\md{i-j}) = \frac{ \moy{b^{\dag}_i b_j}}{\sqrt{n_i n_j}} \;.
\end{equation}
The motivation for this renormalization stems from the observation
that the density-phase expression of the boson creation operator is
$b_i = \sqrt{\rho(x_i)} e^{-i\theta(x_i)}$, and the fact that the
correlator which features superfluid properties in bosonization is
$\moy{e^{i\theta(x_i)} e^{-i\theta(x_j)}}$. For a translationally
invariant model, both definitions only differ by a constant factor.
Since there is no translational invariance, one must likewise average
correlations over all couples of points with the same distance $x =
\md{i-j}$ to obtain a smooth behavior for this correlation. A typical
plot is given in Fig.~\ref{fig:fit-example}. In the case of the BG,
ICDW or MI phases, the Green's function decays exponentially $G(x)
\propto e^{-x/\xi}$. In the Mott phase, the correlation length $\xi$
goes as the inverse one-particle gap $\xi \sim 1/\Delta_c$. An effective
correlation length can also be computed on a finite system
using~\cite{Kuhner2000}:
\begin{equation}
\label{eq:correlation-length}
\xi^2(L) = \frac{ \sum_x x^2 G(x)}{\sum_x G(x)}\;.
\end{equation}
This gives a correct estimate of the correlation length for the
localized phases in the thermodynamical limit up to a factor
$\sqrt{2}$. A divergence of $\xi$ with $L$ signals a superfluid state
in which the asymptotic behavior of the Green's function is algebraic
with an exponent controlled only by the parameter $K$
\begin{equation}
 G(x)  \propto \frac 1 {\displaystyle x^{ 1/2K}}\;.
\end{equation}
This allows for the evaluation of $K$ by using an accurate fitting
procedure on a finite system with open boundary conditions. This is
briefly described in Appendix~\ref{sec:fitting}.

Characterizing the Bose condensation of the cloud is often done by
looking at the condensate fraction $f_c$. It is usually computed on
finite clusters as the largest eigenvalue of the matrix
$\rho_{ij} = \moy{b^{\dag}_j b_i}$. No average over sites nor
normalization by the local density is performed here. The largest
possible value $f_c$ can reach is the number of bosons $N$. In the limit of
HCB, quasi-condensation results in the scaling $f_c(N) \propto
\sqrt{N}$. A finite $f_c$ is a feature of either the BG, the ICDW or
the MI phase. Experimentally, time-of-flight measurements are related 
to the Fourier transform of $\rho_{ij}$, namely
\begin{equation}
\label{eq:nk}
n(k) = \frac 1 L \sum_{lm} e^{ik(l-m)} \rho_{lm} \;.
\end{equation}
Coherence of the quantum gas is deduced from the appearance of a narrow
central peak $n(k=0)$.

\subsection{Localization of free bosons}

\begin{figure}[t]
\includegraphics[height=0.65\columnwidth,clip]{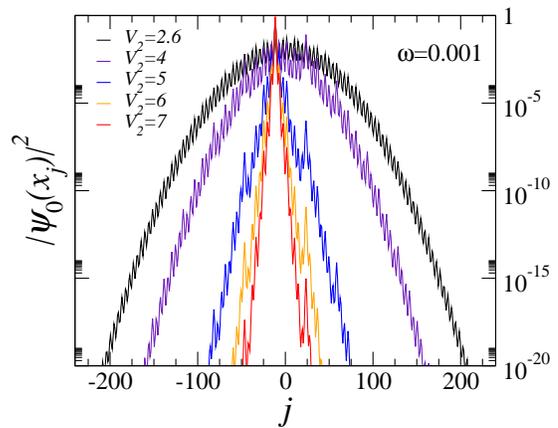}
\caption{(Color online) Ground-state wave function of the lattice
 model (\ref{eq:hamiltonian}) with a smooth trap but no interaction
 for increasing perturbing potential $V_2$. There is a crossover from
 a gaussian wave function (log scale) to an exponential one. Note
 that the maximum of the wave function in the presence of strong $V_2$ is
 not centered at the middle of the trap.}
\label{fig:free-wf}
\end{figure}

\begin{figure}[t]
\includegraphics[height=0.65\columnwidth,clip]{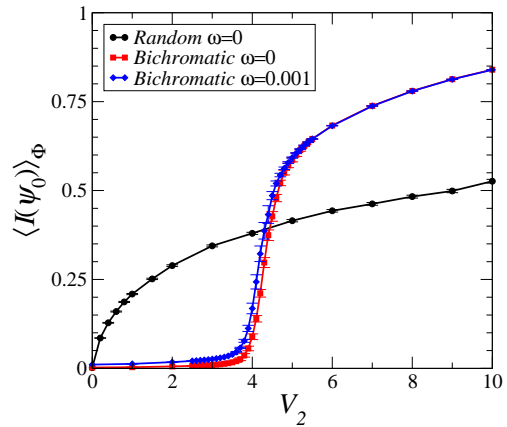}
\caption{(Color online) The averaged inverse participation ratio
 $\moy{I(\psi_0)}_{\phi}$ as a function of $V_2$ for a bichromatic
 lattice and a random box distribution. There is a sharp transition
 at $V_2^c=4$ in the thermodynamical limit ($L=420$ for $\omega=0$).
 A similar sharp transition is also found for the system with a
 smooth trap.}
\label{fig:free-ipr}
\end{figure}

We start with the simplest situation of free bosons, the $U=0$ limit,
in which all bosons lie in the ground-state single particle wave
function $\ket{\psi_0}$.  In Fig.~1 of Ref.~\onlinecite{Lye2007}, the
structure of the trapped wave function is obtained from the
Gross-Pitaevskii equation. Similar results are found here for the
\emph{lattice} model (\ref{eq:hamiltonian}) as shown in
Fig.~\ref{fig:free-wf} which displays the qualitative change of shape
from a gaussian to an exponential structure. In order to quantify the
localization transition of a \emph{single-particle} wave function
$\ket{\psi}$, one can use the inverse participation ratio which is
usually defined as
\begin{equation}
\label{eq:ipr-wf}
I(\psi) = \sum_j \md{\inter{\psi}{j}}^4\;.
\end{equation}
$\ket{j}$ is the state at site $j$ in the real space basis. In the
thermodynamical limit, $I(\psi)$ goes to zero for a delocalized state
with a typical scaling $1/L$ or $\sqrt{\omega}$ for respectively a
non-trapped and a trapped system, while it remains constant for a
localized wave function. Based on an exact duality transformation of
the one-particle Schr\"{o}dinger equation~(\ref{eq:free-hamiltonian}),
the localization of the wave function has been conjectured by Aubry
and Andr\'{e}~\cite{Aubry1980} to happen at the critical value
$V_2^c=4$. This conjecture is illustrated in Fig.~\ref{fig:free-ipr}
which displays $\moy{I(\psi_0)}_{\phi}$ as a function of $V_2$. In
comparison with the RBD evolution, the bichromatic set-up displays a
sharp transition even for a finite trap frequency provided it is small
enough.  For the RBD, localization occurs as soon as $V_2 \neq
0$~\cite{Abrahams1979} with a typical scaling $I(\psi_0) \sim
\sqrt{V_2}$.

\subsection{Localization of hard-core bosons}

\begin{figure}[t]
\includegraphics[height=0.6\columnwidth,clip]{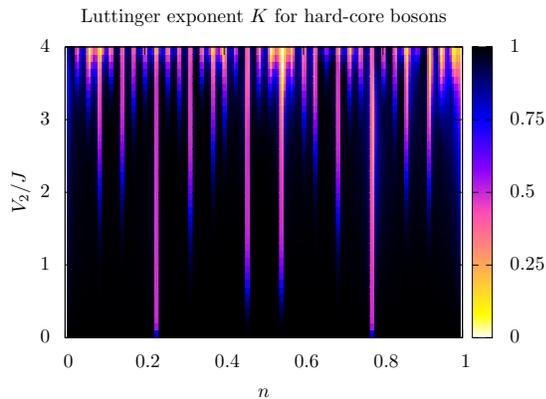}
\caption{(Color online) The Luttinger parameter $K$ for hard-core
 bosons as a function of $V_2$ and the density $n$ on a \emph{finite}
 system with $L=175$ for an irrational $r=0.7714\ldots$. Strictly
 speaking, $K$ should be equal to 1 in all superfluid phases and 0 in
 the gapped phases. Up to finite size effects, vertical bands
 reveal the successive openings of gaps as $V_2$ is increased.}
\label{fig:K4HCB}
\end{figure}

We have seen that plateaus emerge in the $n(\mu)$ curve as soon as
$V_2$ is turned on and that the spectrum is point-like above
$V_2^c=4$. The extension of the wave functions is related to the
nature of the energy spectrum and it was shown~\cite{Aubry1980,
Simon1982} that all wave-functions are extended below $V_2^c$
while they are all localized above. Consequently, we expect the HCB to
localize above $V_2^c$, whatever the density. Below $V_2^c$, HCB can
be either in a SF or in an ICDW state. To illustrate this situation,
we plot in Fig.~\ref{fig:K4HCB} the behavior of the Luttinger exponent
$K$ as a function of $V_2$ and the density $n$. It nicely shows that
$K=1$ in superfluid phases as expected for HCB but vanishes (up to
finite size effects) for the densities corresponding to the ICDW
phases, the main ones being located at $n=r$ and $1-r$. Many gaps
develop as the critical point is approached and the shrinking of the
bands renders the low-energy approximation and calculation of $K$
difficult close to this point.

As a partial conclusion, the two limiting cases $U=0$ and $U=\infty$
of the $(U,V_2)$ phase diagrams of Fig.~\ref{fig:phasediagrams} can be
summed up as follow: (i) for a generic density $n$ (meaning that it
does not correspond to an ICDW plateau) and also for the $U=0$ limit
whatever the density, the system remains superfluid for $V_2 < 4$ and
localizes in a BG phase for $V_2 > 4$ with a correlation length which
behaves according to $\xi^{-1} \sim \ln(V_2/4)$~\cite{Aubry1980,
Simon1982, Thouless1983}, (ii) for a density close to a plateau
phase (for instance $n=r$ or $1-r$) and $U=\infty$, there is a
transition towards an ICDW phase for a critical value of $V_2$ which
is smaller than 4 (and precisely equal to 0 for $n=r$ or $1-r$), (iii) for the
commensurate integer density $n=1$ and $U=\infty$, the system remains
in the MI phase ground-state for any $V_2$.

\subsection{The superfluid -- Bose glass transition for soft-core bosons}

\begin{figure}[t]
\includegraphics[width=0.95\columnwidth,clip]{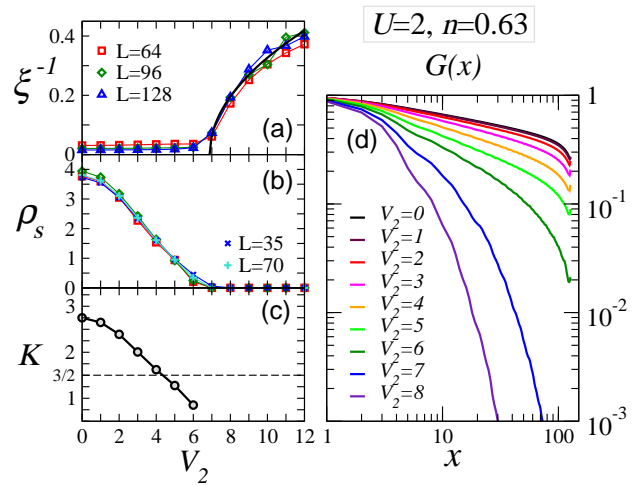}
\caption{(Color online) \emph{Superfluid-Bose glass localization
    transition.} The parameters $U=2$ and $n=0.63$ are chosen such
  that the system is in the superfluid phase at $V_2=0$. (a) The
  inverse correlation length scales roughly as $\sqrt{V_2-V_2^c}$
  where $V_2^c = 6.9$ is much larger than the non-interacting critical
  value $V_2^c = 4$. Below $V_2^c$, we have the scaling $\xi \sim L$
  typical of a superfluid state. (b) $\rho_s$ provides an independent
  determination of the transition. (c) Evolution of the Luttinger
  exponent $K$ as a function of $V_2$. The dashed line displays the
  RBD result $K_c=3/2$. (d) Averaged one-particle density-matrix
  $G(x)$ for increasing $V_2$ showing the transition from an
  algebraic to an exponential decay.}
\label{fig:CorrTransition}
\end{figure}

We address here the direct transition from the SF to the BG phase
which occurs for a generic density by increasing the strength of the
potential $V_2$. Fig.~\ref{fig:CorrTransition} provides the evolution
of the Green's function $G(x)$ showing the localization transition.
First, a finite ``disorder'' strength with a critical value $V_2^c
\simeq 6.9$ is necessary to obtain exponentially decaying
correlations. This value is larger than the $U=0$ and $U=\infty$
limits; interactions have a delocalization effect on the BG phase
similarly to the RBD box results. Qualitatively, this can be
understood by starting from the localized state. There, the condensate
is fragmented into pieces. Repulsive interactions will make the
condensate fragments inflate and, by doing so, will help make them
overlap and build coherence. For bosons, interactions thus helps
delocalization.  Interestingly, computing the Luttinger exponent from
the correlations shows that the critical value $K_c$ at the transition
is smaller than the RBD result $3/2$. The scaling properties of the
transition thus differ from the standard SF-BG transition. Finding a
$K_c$ smaller than the RBD result for the Harper potential is well
compatible with the analytical finding for $K$ found for the Fibonacci
potential in Refs.~\onlinecite{Vidal1999}.

\begin{figure*}[t]
\centering
\includegraphics[width=0.95\textwidth,clip]{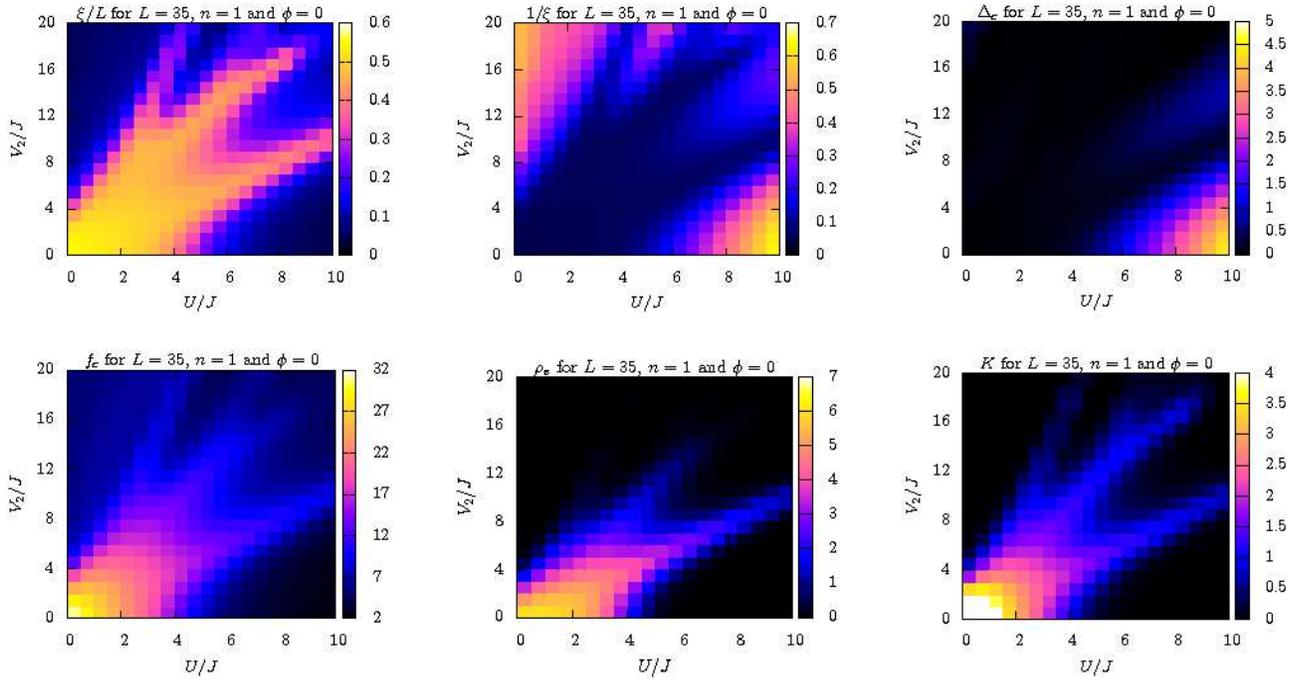}
\caption{(Color online) \emph{Phase diagram for $n=1$.} Observables
 are computed with DMRG for a system with $L=35$ and fixed
 phase-shift $\phi=0$ as a function of $U$ and $V_2$. The $V_2=0$
 line shows the Mott transition at $U_c=3.3$ while the $U=0$ line
 shows the free boson localization transition around $V_2^c=4$.  The
 Mott insulating phase gets qualitatively delocalized as $V_2$
 increases for $U$ not too large. Increasing $U$ delocalizes the BG
 phase if $V_2$ is not too large.}
\label{fig:MapN1}
\end{figure*}

To proceed with the discussion of the competition between interactions
and the disordered potential, we compute with DMRG the phase diagrams
of the system in the three generic cases $n=1$ (competition between
the SF, Mott and BG phases), $n \simeq r$ (competition between the SF,
ICDW and BG phases), and lastly $n=0.5$ (competition between the SF
and BG phases only). The summary of the phase diagrams is given in
Fig.~\ref{fig:phasediagrams}.

\subsection{Phase diagram at $n=1$}

All observables relevant for the construction of the phase diagram as
a function of the interaction $U$ and the potential depth $V_2$ are
reported in Fig.~\ref{fig:MapN1} for the integer density $n=1$. The
Mott phase is characterized by a finite gap $\Delta_c \sim 1/\xi$, a
zero SF density and a finite and small (of order unity) condensate
fraction. It emerges at the bottom right corner above the critical
value~\cite{Kuhner2000} $U_c \simeq 3.3$ for $V_2=0$. We observe that
$U_c$ increases with $V_2$ as for the RBD, meaning that $V_2$
destabilizes the Mott phase. One can understand from a simple local
on-site energies argument: the disorder reduces the minimum
one-particle energy gap in the atomic limit. The BG phase exhibits
exponentially decaying correlations, a zero SF density and a
non-diverging condensate fraction but no gap. It emerges for large
$V_2$ region of the $V_2>U$ half-part of the phase diagram. Note that
the BG has a condensate fraction $f_c$ that is slightly larger than
for the MI phase, qualitatively due to the fact that coherence should
remain significant over the typical length scale of the wells, namely
$1/(1-r)$. The SF phase has a finite SF density but no gap and
algebraic correlations.  It generically emerges at low $U$ and low
$V_2$ and surprisingly extends into a hand print like pattern. A very
small (compared to $U$ and $V_2$) one-particle gap is observed for
large $U$ and large $V_2$ but we cannot conclude whether it is a
finite size effect or not (see Fig.~\ref{fig:Reentrance}).

\begin{figure*}[t]
\includegraphics[width=\textwidth,clip]{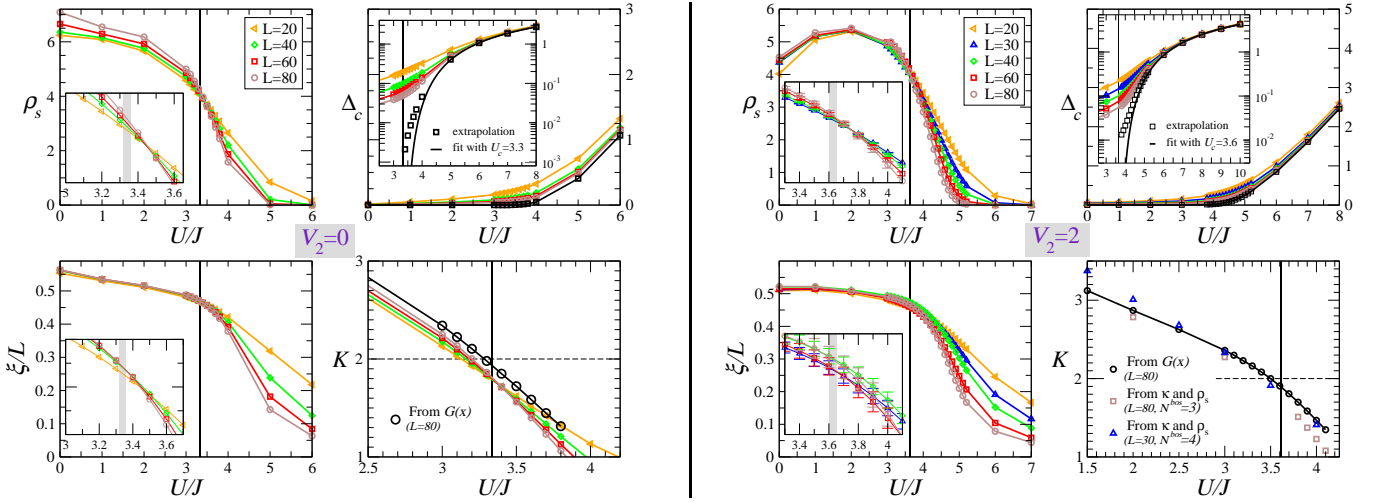}
\caption{(Color online) \emph{Superfluid-Mott insulator transition for
    $n=1$.} Cuts along the $U$-axis for $V_2=0$ and 2. \textbf{Left:}
  data giving similar results to those of Ref.~\onlinecite{Kuhner2000}
  (the vertical bar being $U_c \simeq 3.33 \pm 0.1$). Data are
  computed with $N^{\text{bos}}=4$ and the Luttinger exponent $K$ is
  determined using either $\rho_s$ and $\kappa$ or $G(x)$ (see
  Sec.~\ref{sec:observables}). Scaling of $\rho_s$, $\xi/L$ and the
  criteria $K_c=2$ gives the same critical point within error bars.
  The scaling of the condensate fraction $f_c$ (not shown) is not
  simple at the transition and the HCB scaling $f_c \propto \sqrt{N}$
  does not hold. \textbf{Right:} the same observables for $V_2 = 2$
  and $N^{\text{bos}}=3$ (this cutoff induces a non-physical decrease
  of the superfluidity at small $U$, but does not affect the
  transition as seen, for instance, from the behavior of the Luttinger
  parameter $K$ for $N^{\text{bos}}=4$). Insets show scaling behavior
  for $\rho_s$ and $\xi/L$ after averaging over several different
  $\phi$. A critical point $U_c \simeq 3.6 \pm 0.1$ is found which
  corresponds to $K_c \simeq 2$.  Furthermore, the one-particle gap
  (not averaged over $\phi$) is best fitted by a Kosterlitz-Thouless
  opening (we fixed $U_c=3.6$ for the fit). Up to numerical precision,
  we infer from these results that there is a direct transition
  between the SF and the MI phases, with no intervening BG phase.}
\label{fig:XiCutAll}
\end{figure*}

\emph{Superfluid--Mott transition and intervening Bose-glass phase} --
An important question is whether the SF and Mott touch each other at
small but finite $V_2$. In other words, is there always an intervening
BG phase between the SF and the MI phases as for the
RBD~\cite{Prokofev1998, Rapsch1999}? For the bichromatic potential, we
however have reasons to think that small $V_2$ might not be as
relevant as for true disorder since a large critical value exists for
both hard-core and free bosons. To address this issue numerically, we
have compared the scaling of the most relevant observables for the
known case $V_2=0$ and for $V_2=2$ (see Fig.~\ref{fig:XiCutAll}). When
$V_2 = 0$, the SF-MI transition is of the Kosterlitz-Thouless type
leading to an opening of the one-particle gap $\Delta_c \propto
\exp(-A/\sqrt{U-U_c})$ above the critical value $U_c$, with $A$ a
constant. Such an opening gives a good fit to the extrapolated data
(see Fig.~\ref{fig:XiCutAll}) but does not precisely give $U_c$.
Finding $U_c$ is rather achieved by using the weak-coupling RG result
$K_c=2$ for the KT transition. Fig.~\ref{fig:XiCutAll} shows that $U_c
\simeq 3.3 \pm 0.1$ for $V_2 = 0$ in agreement with results of
Ref.~\onlinecite{Kuhner2000}. Within error bars, the scalings of the
superfluid stiffness $\rho_s$ and correlation length also agree with
this critical point. Note that because of the very slow opening of the
one-particle gap in a KT transition, the correlation length and
superfluid density show much smoother finite size effects than for the
SF--BG transition illustrated in Fig.~\ref{fig:CorrTransition}.  For
$V_2 = 2$, if a BG is present in between the SF and MI phase, the
one-particle gap $\Delta_c$ should open after the superfluid stiffness
scales to zero. Up to numerical accuracy, data are consistent with a
direct SF-MI transition of the KT type with a slightly larger critical
interaction $U_c \simeq 3.6 \pm 0.1$. We observed that averaging over
$\phi$ is needed to ensure a good crossing of the scaling curves (see
insets of Fig.~\ref{fig:XiCutAll}). Note that for the RBD situation,
$V_2=2$ would correspond to a disorder amplitude $\Delta=1$ in
Ref.~\onlinecite{Rapsch1999} (or $\Delta = 0.5$ in
Ref.~\onlinecite{Prokofev1998}) for which the BG phase already has a
significant width. Interestingly, the $n=1$ phase diagram has a
similar shape as the one with a commensurate potential with
$r=1/2$~\cite{Rousseau2006}. In this case, there is at large $V_2$ a
charge density wave phase is a gapped with two particles each two
sites for which $K_c=1/2$. A direct SF-MI transition is found because
the term (\ref{eq:bichropot}) will not be relevant for small $V_2$.
Our results suggest that in the incommensurate case, the potential
remains irrelevant as well.

\begin{figure}[t]
\includegraphics[width=0.95\columnwidth,clip]{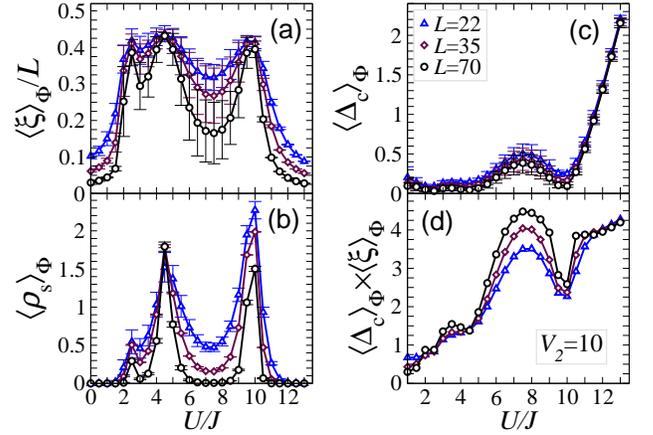}
\caption{(Color online) \emph{Reentrances of the SF phase with
    increasing interactions.} Cut at $V_2=10$ in the phase diagram of
  Fig.~\ref{fig:MapN1}. Error bars are related to the average over the
  phase-shift $\phi$ (about twenty samples). The scalings of the
  correlation length (a) and of the superfluid density (b) suggest two
  reentrances of the SF phase as $U$ is increased. In between, a
  localized phase is found and a MI phase is obtained at large $U$
  according to the gap (c). In the intermediate BG phase (for $U\simeq
  6$--9), a small gap is found, but for the chosen sizes $L=22,35$ and
  70, the scaling $\xi \sim 1/\Delta$ shown in (d) seems to be
  satisfied only in the MI phase, meaning that strong finite size
  effects are present.}
\label{fig:Reentrance}
\end{figure}

\begin{figure*}[!ht]
\centering
\includegraphics[width=0.95\textwidth,clip]{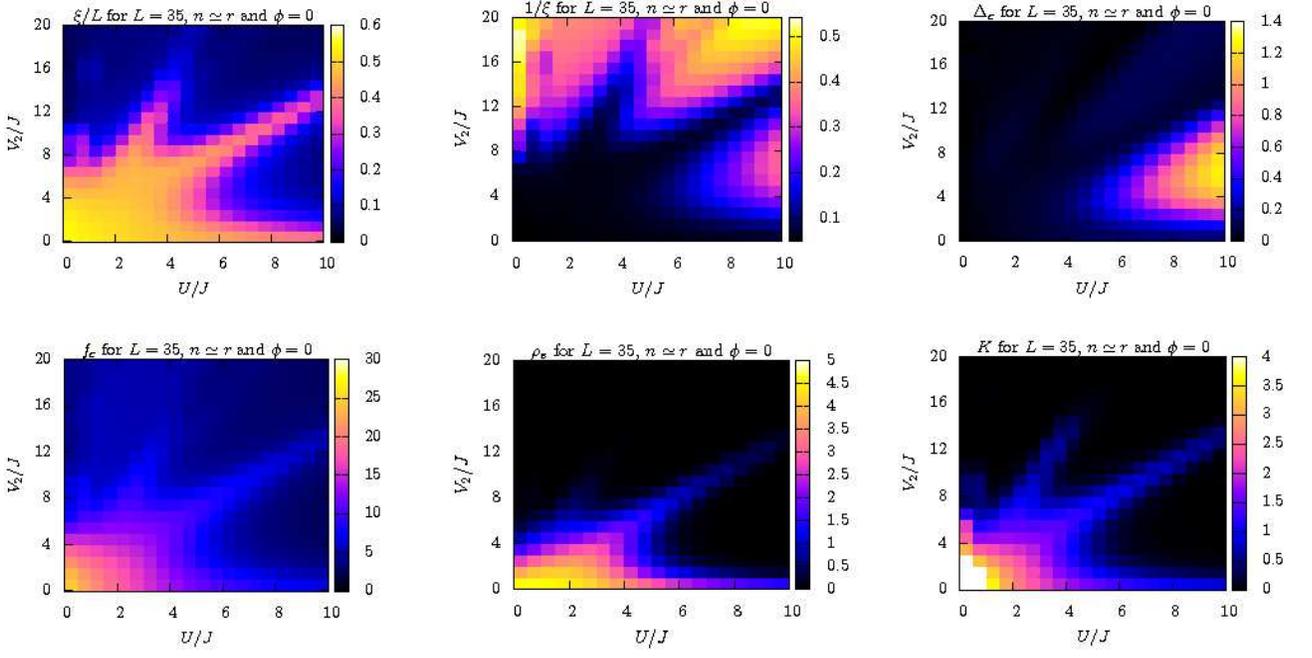}
\caption{(Color online) \emph{Phase diagram for $n \simeq r$}.
 Observables are computed on a system with fixed size $L=35$ and
 fixed phase-shift $\phi=0$. A large ICDW phase emerges at large $U$.
 A finite $V_2$ is required to stabilize it. The SF phase extends
 along the $V_2=0$ contrary to the phase diagram with $n=1$.}
\label{fig:MapNR}
\end{figure*}

\emph{Superfluid--Bose Glass transition} -- We now turn to the
discussion of the contour of the SF-BG transition which displays a
``hand print'' pattern. First, contrary to the RBD, the BG phase
emerges only above $V_2^c=4$ and for much larger values for small $U$.
Secondly, $V_2^c$ increases with $U$ at small $U$ which is similar to
the delocalization by interactions observed in the RBD case.
Similarly to what was found in Fig.~\ref{fig:CorrTransition}, the
inverse correlation length has a power-law behavior above the critical
point with a Luttinger parameter smaller than $3/2$. The convexity of
the SF phase contours changes contrary to the RBD phase diagram,
leading to this hand print pattern. To understand if these reentrances
of the SF phase inside the BG phase are not a finite size effect and
remain after averaging over $\phi$, we show the averaged
$\moy{\rho_s}_{\phi}$ and $\moy{\xi}_{\phi}$ for various system sizes
in Fig.~\ref{fig:Reentrance}. The behavior of $\xi$ and $\rho_s$
suggests two reentrances of the SF phase and in particular, a sharp
but clear one close to the transition to the MI phase. The $U=V_2$
line corresponds to the ``transition'' between the MI and BG phases in
the atomic limit. It gives a rough estimate for the extension of a SF
phase at large $U$ and $V_2$ which does not occur for the RBD
situation. We expect that the SF phase vanishes for large $U$ and that
there is a direct MI-BG transition around the $U=V_2$ line. A similar
emergence of the superfluid phase around the atomic limit was found in
Ref.~\onlinecite{Rousseau2006} for the case of a commensurate
potential where the SF phase competes a CDW and a MI phase. In
Fig.~\ref{fig:Reentrance}, the intermediate localized phase between
the two SF reentrances displays a small gap. This phase could have a
finite gap but we observe that if so, it cannot be distinguished from
finite size effects. Comparing again with the commensurate
case~\cite{Rousseau2006}, the main difference (apart from the finite
gap) is the extension of the SF phase along the $U=0$ line. For the
Harper model, there is no such extension because of the localization
of the single-particle wave-function.

\subsection{Phase diagram close to the density $n=r$}

A density which satisfies the criteria $n \simeq r$ allows for the
realization of an ICDW phase which competes with the SF and BG phases. The
$(U,V_2)$ map of the observables is given in Fig.~\ref{fig:MapNR}.
The ICDW phase has a finite gap and exponentially decaying
correlations as the MI phase. Similar qualitative features are found
with the ICDW phase replacing the MI phase. However, a finite $V_2$ is
of course required to stabilize the ICDW phase contrary to the MI
phase. Secondly, a finite $V_2$ is needed to stabilize the BG phase.
As a consequence, the SF phase extends to large $U$ close to the
$V_2=0$ line. As discussed in section~\ref{sec:plateaus}, the ICDW is
a new feature compared with the RBD phase diagram given in
Refs.~\onlinecite{Prokofev1998, Rapsch1999} for $n=0.5$. Similar
reentrances of the SF phase into the BG phase are found at fixed $V_2$
and increasing $U$. The $U=\infty$ line of the phase diagram would
give an ICDW phase everywhere except for $V_2=0$ since the $n=r$
plateau occurs as soon as $V_2$ is finite in the HCB limit.

\subsection{Phase diagram for a generic density $n=1/2$}

Lastly, the phase diagram for a generic density $n=1/2$ has been
computed to discuss only the competition between the SF and BG phases
(data not shown, see phase diagram in Fig.~\ref{fig:phasediagrams} for
results). We must note that, ICDW plateaus can however appear for
generic density in a region where $U \sim V_2$, as we show in
Fig.~\ref{fig:N_of_mu_U6V8} for the particular choice of parameters
$V_2=8$ and $U=6$. In this case, the ICDW phase would have a finite
domain in the $(U,V_2)$ map (contrary to the $n=r$ phase diagram),
because the ICDW phase is not realized in the HCB limit. The
observables suggest that $\Delta_c$ remains zero in the whole
parameter range, while the BG is bounded by the $V_2=4$ line and the
SF phase slightly extends inside the BG phase for small $U$. However,
critical values $V_2^c$ for the SF-BG transition are found to be
smaller than for $n=r$, themselves smaller than for $n=1$. The same
qualitative argument stating that the lower the density, the closer
the physics is to the HCB can be put forward. The SF region extends
with the density of the system. This observation will be now more
precisely discussed.

\subsection{Delocalization via increasing the density of bosons}
\label{sec:delocalization}

\begin{figure*}[t]
\includegraphics[width=0.95\textwidth,clip]{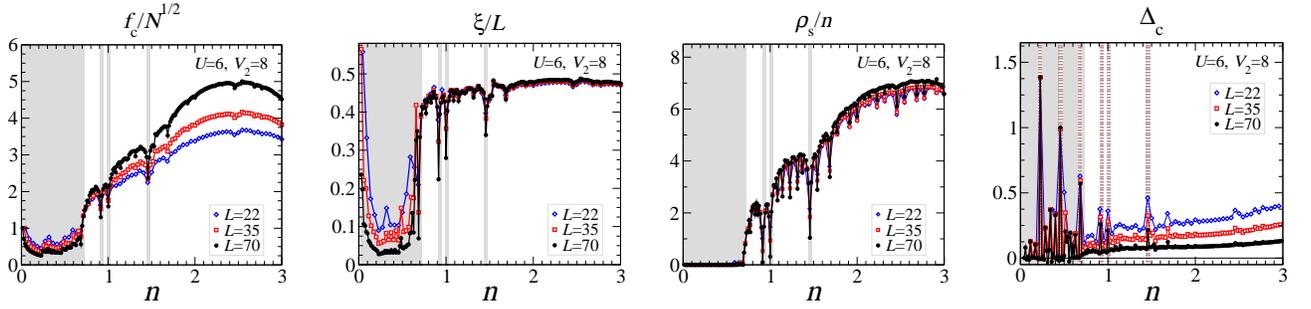}
\caption{(Color online) \emph{Delocalization by increasing the number
    of particles.} Observables for increasing density when $V_2 \simeq
  U$ (same parameters as in Fig.~\ref{fig:N_of_mu_U6V8}). There is a
  delocalization transition with increasing density. A few ICDW phases
  can be seen at intermediate fillings (dashed brown vertical lines in
  the r.h.s. figure showing the one=particle gap $\Delta_c$). Grey
  areas denote the localized regions (either BG or ICDW). The
  weakening of the superfluidity at large density might be an artefact
  of the cut-off in the number of bosons kept per site which is here
  $N^{\text{bos}}=4$.}
\label{fig:XiCutN}
\end{figure*}

An complementary approach to these $(U,V_2)$ phase diagrams at fixed
density is to keep $U$ and $V_2$ constant and to look at the
observables as a function of the density $n$. From dipole oscillations
measurements, Lye~\emph{et al.}~\cite{Lye2007} observed a
delocalization transition by increasing the number of particles. We
now address the non-trivial case of $U \sim V_2$ by setting $U=6$ and
$V_2 = 8$ corresponding to the parameters of the $n(\mu)$ curve of
Fig.~\ref{fig:N_of_mu_U6V8}. Results for the same observables as for
the phase diagrams are plotted in Fig.~\ref{fig:XiCutN}. We found
transitions between the three different phases BG, ICDW and SF. At low
densities, double occupation for bosons is strongly suppressed because
of the finite $U$.  Consequently, the behavior is qualitatively the
one HCB would have: $V_2$ being larger than 4, localization exists at
low densities. The superfluid density, correlation length and
one-particle gap confirm the presence of the BG phase. At large
densities, a SF emerges which is something well-known without disorder
because the lobes of the Mott phases shrink at large
densities~\cite{Kuhner2000}.  In addition, the disordered potential
has a tendency to reduce the size of the Mott phases as we have seen.
Very qualitatively, some particles fill the wells of the disorder
potential so that the remaining ones feel a smoother effective
potential allowing for a gain in kinetic energy leading to
superfluidity. This behavior for an irrational $r$ is qualitatively
similar to what was observed for a rational $r$ (see Fig.~23 of
Ref.~\onlinecite{Rousseau2006}) except that no BG, but a ``weakly
superfluid'' phase is realized in this latter case. Besides this sharp
BG-SF transition, peaks in the one-particle gap $\Delta_c$ uncover the
presence of ICDW phases within both the BG and the SF phases. These
phases naturally correspond to the plateaus in
Fig.~\ref{fig:N_of_mu_U6V8}.

\section{Probing the Bose-glass phase with out-of-equilibrium
 dynamics}
\label{sec:dynamics}

\begin{figure}[b]
\includegraphics[width=0.85\columnwidth,clip]{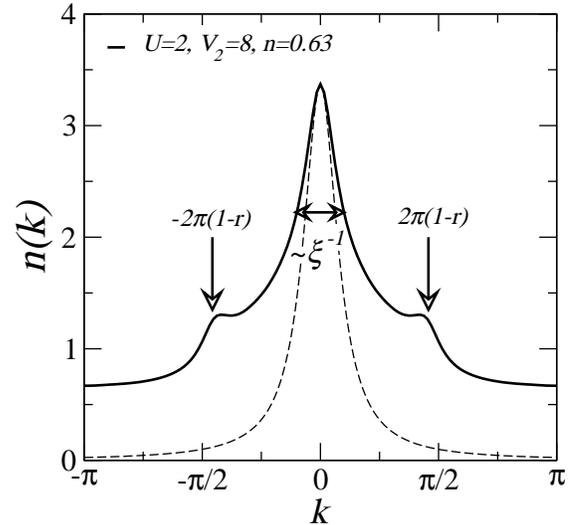}
\caption{Typical $n(k)$ in the Bose-glass phase of
  Fig.~\ref{fig:CorrTransition} in a system without a trap. Satellites
  peaks at wave vector $2\pi (1-r)$ are visible if the disorder is not
  too strong. The width of the middle peak typically gives the inverse
  correlation length (a Lorentzian of width $\xi$ computed from
  Eq.~(\ref{eq:correlation-length}) is given in dashed lines for a
  qualitative comparison).}
\label{fig:Nk}
\end{figure}

\subsection{Static observables}

The question of probing experimentally the BG phase with respect to
the other possible phases is particularly important. First, the
simplest observable obtained after time-of-flight measurements is
related to the momentum distribution of the atoms $n(k)$. This measure
helps distinguish between coherent and incoherent phases by looking,
in particular, at the $k=0$ peak. A sharp and high peak is the
signature of a coherent phase, the superfluid phase. Because of
short-range correlations, both the MI and the BG phases will give a
much smaller peak broadened with a typical width of $\xi^{-1}$.
Fig.~\ref{fig:Nk} displays $n(k)$ in the BG at small $U$. In addition
to the central peak, satellites peaks at $k=\pm 2\pi(1-r)$ emerge as a
signature of the underlying superlattice. However, in the experiment
pictures, the Wannier envelope and the broadening of the peaks due to
scattering events during the time-of-flight will change the observed
shape. It is expected that the additional satellite peaks are too
small to be experimentally resolved and are washed out if either $V_2$
and/or $U$ are too large. Thus, $n(k)$ can only be used to distinguish
the superfluid from the Bose-glass or Mott insulating phases. However,
it would not help distinguish the MI from the BG phase.
Refs.~\onlinecite{Roth2003, Roscilde2007} found a similar behavior
and, in the second reference, a non-monotonic evolution of the central
peak $n(k=0)$ with increasing $V_2$ has been established. The
reinforcement of the superfluidity upon increasing $V_2$ at fixed $U$
in a trapped cloud must be reminiscent of the MI-SF-BG transitions of
the phase diagrams of Fig.~\ref{fig:phasediagrams}. Noise
correlations~\cite{Altman2004} were proposed~\cite{Rey2006,
  Roscilde2007} as a possible probe for the BG phase and measured in
Ref.~\onlinecite{Guarrera2008}. However, this observable catches the
fact that density correlations reveal the presence of the underlying
superlattice~\cite{Rey2006, Roscilde2007} but not the gapless nature
of the excitations~\cite{Roscilde2007}. It is therefore necessary to
look for additional evidence of localization.

\subsection{Expansion in the lattice potential}

\begin{figure}[t]
\centering
\includegraphics[width=\columnwidth,clip]{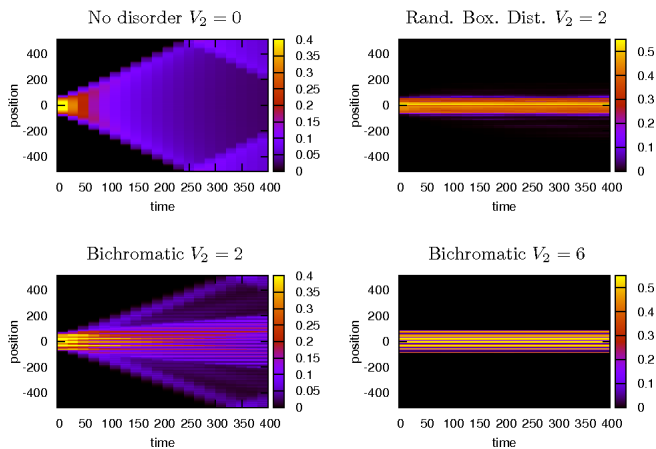}
\caption{(Color online) \emph{Expansion of HCB condensates}. At $t=0$,
  the system is in the ground-state of the hamiltonian with the trap
  plus ``disorder'' potentials using a fixed chemical potential
  $\mu=0$ (about 51 particles) and a trap frequency $\omega=0.03$. At
  $t>0$, the trap confinement is switched off abruptly. Figures show
  the evolution of the local density profile as a function of time. An
  expansion is observed for systems with $V_2=0$ (without disorder)
  and $V_2=2$, but not for a random potential and a bichromatic
  potential with $V_2=6 > V_2^c$. For the RBD, $V_2$ defines the width
  of the box distribution. When there is an expansion, reflections on
  the boundaries of the box in which the condensate expands can be
  seen.}
\label{fig:dyn-expansion}
\end{figure}

As often done in experiments~\cite{Clement2005, Fort2005, Lye2005,
  Clement2006, Clement2008, Chen2007, Lye2007}, transport measurements
are a better fashion to probe localization. In order to show the
existence of a critical point for the localization, we propose to look
at the expansion of the cloud when the trap is
released~\cite{Rigol2004a, Fort2005, Clement2005, Rigol2005,
  Clement2006, Clement2008, Sanchez-Palencia2007}. Observing the
expansion in the optical lattice is a particularly appealing
experiment as the hamiltonian governing the dynamics is the one of the
bulk system (with $\omega =0$) for which we have computed the
equilibrium phase diagram. The confinement is used here to prepare an
out-of-equilibrium state for this hamiltonian.

For the sake of clarity, we first discuss the expansion of HCB. The
spreading has been studied before in the HCB limit for homogeneous
lattices~\cite{Rigol2005}, and for soft-core bosons in commensurate
lattices~\cite{Rodriguez2006}.  Fig.~\ref{fig:dyn-expansion} displays
the expansion for free HCB ($V_2=0$), for two bichromatic potential
amplitudes, below ($V_2=2$) and above ($V_2=6$) the equilibrium
critical point, and also a situation with a RBD potential of an
amplitude $V_2=2$. The system is prepared in the ground-state of the
hamiltonian with the confining potential (we chose a trap frequency of
$\omega=0.03$). At $t=0$, the trap frequency is set to zero and the
condensate is free to expand into the lattice. For $V_2=0$, the
expansion of the edges of the condensate is roughly linear, with a
typical velocity $2J$ corresponding to the maximum group velocity (see
below). For the RBD potential, the expansion is inhibited for the
amplitude $V_2=2$.  However, for the bichromatic set-up, the same
potential strength does not prevent the condensate from expanding.
Still, $V_2=6$ induces a localization of the condensate similar to the
one observed for the RBD potential.

\begin{figure}[t]
\centering
\includegraphics[width=0.75\columnwidth,clip]{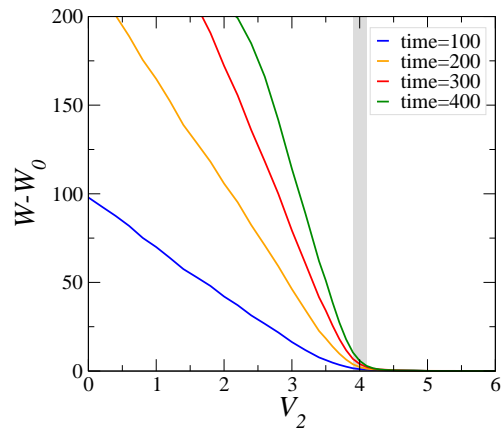}
\caption{(Color online) \emph{Dynamical critical point} -- Width of
  the atomic distribution $W=\sqrt{\overline{n_j(j-j_0)^2}}$ of a HCB
  condensate as a function of $V_2$ for several increasing times (in
  units if the hopping). The width $W_0$ at $t=0$ has been subtracted
  for clarity. Parameters are the same as in
  Fig.~\ref{fig:dyn-expansion}. Up to finite trap frequency effects
  (see text), the dynamical critical point is identical to the
  equilibrium one (at least for HCB).}
\label{fig:dyn-expansion-crit}
\end{figure}

\begin{figure*}[t]
\centering
\includegraphics[width=\textwidth,clip]{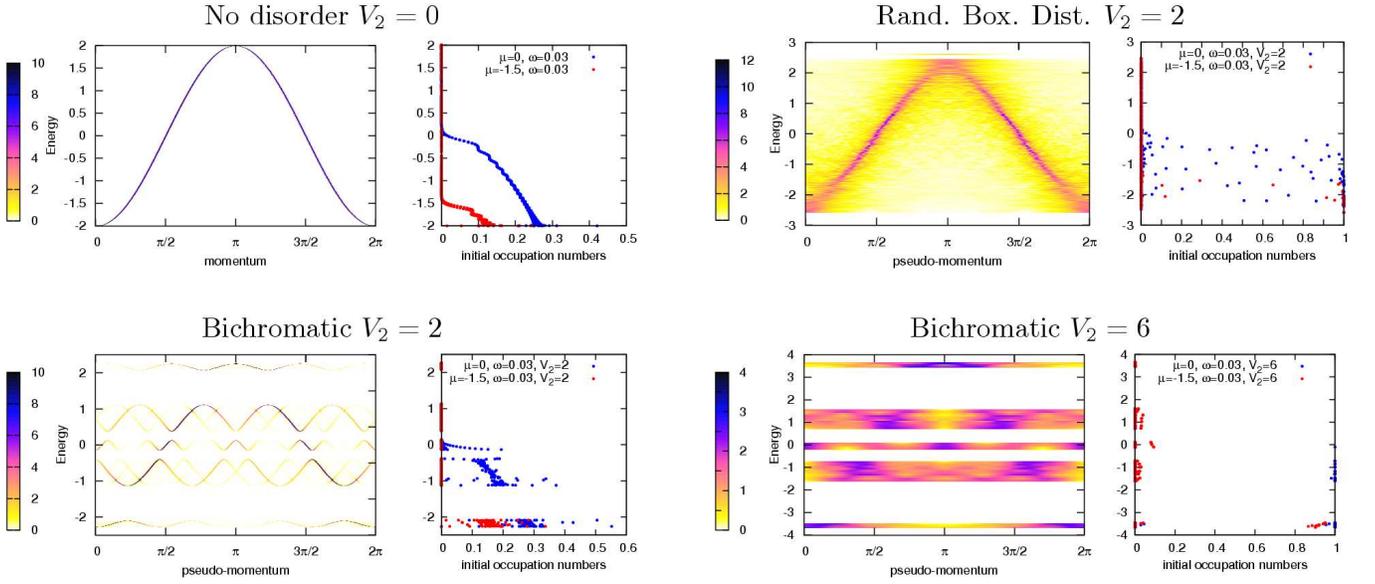}
\caption{(Color online) \emph{Effective dispersion relations for HCB}
  -- the Fourier transform of the single-particle wave functions
  $\md{\psi_k}^2$ is plotted as a function of the single-particle
  energy and the pseudo-momentum $k$. At the right each dispersion
  curve is the initial occupation numbers (from the state prepared
  using the trap confinement) v.s. the same single-particle energies
  for two different chemical potential $\mu =-1.5, 0$. This shows
  which states participate to the expansion. Other parameters are the
  same as for the expansions shown in Fig.~\ref{fig:dyn-expansion}.}
\label{fig:dyn-dispersion}
\end{figure*}

One may ask whether the critical point of the dynamical localization
observed in Fig.~\ref{fig:dyn-expansion} is the same as the
equilibrium one. In the case of HCB, we know that all single-particle
wave-functions are localized above $V_2=4$ (see
Sec.~\ref{sec:hcb-plateaus} and references therein). Consequently, we
expect the dynamical critical point to be identical to the equilibrium
one. To support this statement, we show in
Fig.~\ref{fig:dyn-expansion-crit} the width of the atomic distribution
of the condensate after several times of expansion, as a function of
the ``disorder strength'' $V_2$. The dynamical critical point is found
to be very close to $V_2=4$, within a $\pm 0.1$ window (grey rectangle
in Fig.~\ref{fig:dyn-expansion-crit}).  One observes that, slightly
above $V_2=4$, the condensate still spreads a little bit with time.
This may be understood as a finite trap frequency effect. Indeed, the
edges of the condensate can spread over a few sites if the initial
trap is too steep. In this case, the starting atomic distribution is
too far from the local atomic distribution that is expected (locally)
in the bulk of a non-trapped system, and particles have to be
redistributed. In the limit of vanishing initial trap frequency, we
expect the transition to be sharper. Note that, as the localization of
all single-particle wave-functions in the spectrum does not depend on
$r$ (provided it is irrational), we expect the equality of the
dynamical and equilibrium critical points to hold independently of
$r$.  Fig.~\ref{fig:dyn-expansion-crit} may be interpreted as the
vanishing of all effective group velocities at the critical point (see
below). Thus, the exact critical value $V_2^c=4$ for the localization
could be probed experimentally with this technique for free or hard
core bosons.

Following Ref.~\onlinecite{Diener2001}, a more precise description of
the HCB expansion can be carried out by looking at the one-particle
effective dispersion $\varepsilon(k)$ for HCB. Without translational
symmetry, wave vectors $k$ are not good quantum numbers but looking at
the Fourier transform of the one-particle wave functions $\psi_k \sim
\sum_j e^{ikj} \psi_j$ and plotting $\md{\psi_k}^2$ as a function of
the pseudo-momentum $k$ provides an effective dispersion. The features
of the expansion depend mainly on two properties. First, the group
velocities $v_g(k) = \partial \varepsilon(k)/\partial k$ derived from
the effective dispersion relation convey the typical maximal speed at
which expansion evolves. Second, the expansion also strongly depends
on the initial occupation numbers $n_{\varepsilon}(t=0)$ of the
eigenstates of the hamiltonian with $\omega=0$. This occupation is
plotted together with the dispersion relation as a function of the
``single-particle energy'' $\varepsilon$ in
Fig.~\ref{fig:dyn-dispersion} corresponding to the expansion observed
in Fig.~\ref{fig:dyn-expansion}. For $V_2=2$, for which there is no
localization, the effective relation dispersion displays gaps as we
have seen from section~\ref{sec:plateaus} and well-defined bands with
a shorter periodicity originating from the band foldings induced by
the potential (see Sec.~\ref{sec:plateaus}). Compared with the single
cosine dispersion obtained without disorder, several shifted bands
exist due to Bragg scattering with the potential. One can convince
oneself that opening gaps lowers the maximum possible group velocity.
Thus, compared to a system with no disorder
(Fig.~\ref{fig:dyn-dispersion}, $V_2=0$), the expansion for the
bichromatic potential below $V_2=4$ will always be slower if the
$\varepsilon=0$ state (associated with the maximum group velocity
$2J$) is occupied in the initial state without disorder. This explains
the qualitative features of the situations for which the condensate
expands in Fig.~\ref{fig:dyn-expansion}. When $V_2=0$, the expansion
is slower when the chemical potential is below much 0 (not shown), as
can be guessed from the initial occupation numbers in
Fig.~\ref{fig:dyn-dispersion}.  When $V_2=2$, the structure of the
expansion is rather homogeneous at low chemical potential (not shown)
but becomes inhomogeneous and faster for larger $\mu$ (see
Fig.~\ref{fig:dyn-expansion}). The presence of two different speeds
might stem from populating bands with different maximum group
velocities as can be seen in Fig.~\ref{fig:dyn-dispersion}. For the
bichromatic potential with $V_2=6$, no bands can be distinguished as
the signal does not show well-defined pseudo-momentums. The RBD
potential displays a very different effective dispersion faded by the
disorder, but which still retains the whole feature of the cosine
dispersion without disorder. These two pictures illustrate that the
localization mechanism for the bichromatic and a RBD potential is
qualitatively different: the first one is rather associated with a
band folding mechanism while the second rather corresponds to strongly
scattered single-particle states. In this respect, one can view the
``weakly superfluid'' phase found for commensurate superlattices with
a large $V_2$ in Ref.~\onlinecite{Rousseau2006} as a precursor of the
Bose-glass phase of incommensurate lattices.

How these results can carry to investigate the physics at finite $U$
is an important and challenging question that needs further
investigations going clearly beyond the goal of this paper. Indeed,
from the numerical point of view, the expansion of strongly correlated
soft-core bosons is accessible with time-dependent DMRG only until
times of order $10J$ \cite{Rodriguez2006}, while
Fig.~\ref{fig:dyn-expansion-crit} shows that a reasonable
determination of the out-of-equilibrium critical point requires times
of, at least, $100J$. Thus, the question of the dynamical localization
at finite $U$ of the model (\ref{eq:hamiltonian}) and its relation
with the equilibrium phase diagrams of Fig.~\ref{fig:phasediagrams}
remains an open question. First, the most naive prediction would be to
expect a similar physics than the HCB one to hold, at least for large
enough interations. Further qualitative arguments can be given on the
expansion for intermediate $U$, for which we can use two results from
the equilibrium phase diagrams studied in Sec.~\ref{sec:diagrams}: (i)
the critical values to observe localization are all larger than
$V_2=4$, whatever $U$ or the density, (ii) at small densities, the
physics is essentially equivalent to the one of HCB. First consider a
situation where $V_2 \leq 4$. The starting trapped state is expected
to have regions which can be locally SF or MI but not BG, since there
is no intervening BG phase. Provided $V_2\leq 4$, an expansion is then
expected systematically (whatever $U$ or the total number of
particles), because the edges of the condensate would be in a SF state
(see examples of expansions of strongly-correlated soft-core bosons in
Ref.~\cite{Rodriguez2006}).  For $V_2 > 4$, the situation is more
subtle as, for a given $U$ and $V_2$, the occurence of localization
depends on density in the non-trapped condensate (see for instance
Fig.~\ref{fig:XiCutN}).  The starting state structure is complex and
local-density approximation not necessarily valid~\cite{Roscilde2007}.
Very qualitatively, the edges of the condensate would be in a
localized state while, if the density at the center of the cloud is
large enough, SF or Mott regions could also appear. However, if
expansion there is, the local density will decrease with time. When
the density becomes small enough to neglect interactions, localization
would then be expected since $V_2 > 4$ and one enters the HCB regime.
A possible scenario could thus be a systematic localization after
either a transient regime with expansion, or no transient regime.
However, ascertaining whether the above qualitative arguments could be
spoiled by other effects is difficult. For instance, it is known in
the completely different limit of very weakly interacting bosons
(Gross-Pitaevskii limit) that non-linear effects due to interactions
can lead to some kind of localization even in a purely effective
periodic potential~\cite{Trombettoni2001,Anker2005}. How to go from
such a limit to the relevant one for the Anderson localization in
strongly interacting one dimensional systems is clearly a question
that will need further experimental and theoretical work.

\section{Conclusion}

The Bose-Hubbard model with a quasi-periodic potential was shown to
display a rich phase diagram including a Bose-glass phase (localized
but compressible), and incommensurate charge-density wave phases in
addition to the superfluid and Mott phases. While localization induced
by this random-like potential is found, the underlying mechanism
differs from the RBD situation: the band folding mechanism known
previously for free and hard-core bosons (or fermions) holds for
soft-core bosons, leading to a finite critical value of the
localization transition $V_2^c \geq 4$. The critical values found are
high, possibly sufficiently high to allow for an experimental
demonstration of a localization transition. In this perspective,
static observables give a clear evidence to distinguish between
coherent and localized phases, but their ability to sort the BG from
the (small-$V_2$) MI phase is less obvious. On the contrary, the
expansion of the condensate after switching off the confinement is
proposed to provide a simple and rather clear signal to detect the
localization transition. This was shown explicitly in the case of
hard-core bosons but remains an open question for soft-core bosons.

\acknowledgments

We thank Fabian Heidrich-Meisner, Alexei Kolezhuk, Massimo Inguscio
and Tommaso Roscilde for fruitful discussions. T.B. acknowledges the
Studienstiftung des deutschen Volkes for financial support. This work
was supported in part by the Swiss National Science Foundation under
MaNEP and Division II, DARPA OLE programme, and by the DFG. C.K.
acknowledges support of the RTRA network ``Triangle de la Physique''.

\appendix

\section{Method to fit the bosonic Green's function on finite systems}
\label{sec:fitting}

\begin{figure}[t]
\includegraphics[width=0.7\columnwidth,clip]{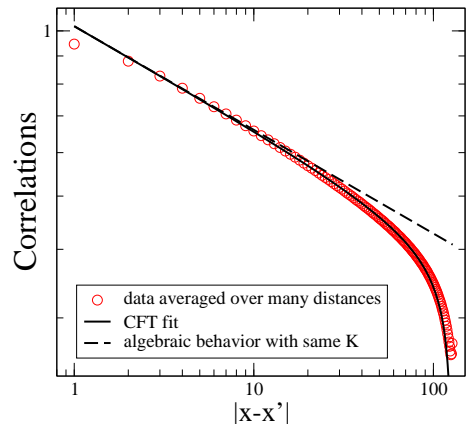}
\caption{(Color online) A typical example of fitting the averaged
 data from DMRG with conformal field theory results to extract the
 Luttinger exponent $K$. System size is $L=128$.}
\label{fig:fit-example}
\end{figure}

We use conformal field theory results~\cite{Cazalilla2004} for a
system of length $L$ with open boundary conditions to fit the bosonic
Green's function defined in Eq.~(\ref{eq:green-function}). In the case
of correlations of the type $\moy{e^{i\theta(x)}e^{-i\theta(x')}}$ one
has :
\begin{equation}
\label{eq:cft-fit}
G(x,x') = A \left[ \frac{ \sqrt{d(2x \vert 2L) d(2x' \vert 2L)} }
{d(x+x' \vert 2L)d(x-x' \vert 2L) } \right]^{\displaystyle \frac{1}{2K} }\,,
\end{equation}
with $K$ the Luttinger parameter, $A$ a constant and $d$ the conformal length 
\begin{equation*}
d(x|L) = \frac{L}{\pi}\left\vert\sin\left(\frac{\pi x}{L}\right)\right\vert\,.
\end{equation*}
Because there is no translational invariance, the correlations depend
on both $x$ and $x'$. Hence, in order to perform a fit, one has to
average Eq.~(\ref{eq:cft-fit}) over the results obtained with fixed
distance $x'-x$. Strictly speaking, formula~(\ref{eq:cft-fit}) is
valid for $1 \ll x, x', \md{x'-x} \ll L$ and we have to remove the
corresponding contributions. Practically, fits are rather good up to
distances comparable with $L$ as one can see in
Fig.~\ref{fig:fit-example} and significantly improves the
determination of $K$ compared with a simple algebraic fit.


\begin{thebibliography}{10}

\bibitem{Anderson1958}
P.~W. Anderson, Phys. Rev. {\bf 109}, 1492 (1958).

\bibitem{Lee1985}
P.~A. Lee and T.~V. Ramakrishnan, Rev. Mod. Phys. {\bf 57}, 287 (1985).

\bibitem{Abrahams1979}
E.~Abrahams, P.~W. Anderson, D.~C. Licciardello, and T.~V. Ramakrishnan, Phys.
  Rev. Lett. {\bf 42}, 673 (1979).

\bibitem{Giamarchi1987}
T.~Giamarchi and H.~J. Schulz, Europhys. Lett. {\bf 3}, 1287 (1987).

\bibitem{Giamarchi1988}
T.~Giamarchi and H.~J. Schulz, Phys. Rev. B {\bf 37}, 325 (1988).

\bibitem{Fisher1989}
M.~P.~A. Fisher, P.~B. Weichman, G.~Grinstein, and D.~S. Fisher, Phys. Rev. B
  {\bf 40}, 546 (1989).

\bibitem{Batrouni1990}
G.~G. Batrouni, R.~T. Scalettar, and G.~T. Zimanyi, Phys. Rev. Lett. {\bf 65},
  1765 (1990).

\bibitem{Scalettar1991}
R.~T. Scalettar, G.~G. Batrouni, and G.~T. Zimanyi, Phys. Rev. Lett. {\bf 66},
  3144 (1991).

\bibitem{Prokofev1998}
N.~V. Prokof'ev and B.~V. Svistunov, Phys. Rev. Lett. {\bf 80}, 4355 (1998).

\bibitem{Rapsch1999}
S.~Rapsch, U.~Schollw\"{o}ck, and W.~Zwerger, Europhys. Lett. {\bf 46}, 559
  (1999).

\bibitem{Bloch2007}
I.~Bloch, J.~Dalibard, and W.~Zwerger, Rev. Mod. Phys. {\bf 80}, 885 (2008).

\bibitem{Lye2005}
J.~E. Lye, L.~Fallani, M.~Modugno, D.~S. Wiersma, C.~Fort, and M.~Inguscio,
  Phys. Rev. Lett. {\bf 95}, 070401 (2005).

\bibitem{Clement2005}
D.~Cl\'{e}ment, A.~F. Var\'{o}n, M.~Hugbart, J.~A. Retter, P.~Bouyer,
  L.~Sanchez-Palencia, D.~M. Gangardt, G.~V. Shlyapnikov, and A.~Aspect, Phys.
  Rev. Lett. {\bf 95}, 170409 (2005).

\bibitem{Fort2005}
C.~Fort, L.~Fallani, V.~Guarrera, J.~E. Lye, M.~Modugno, D.~S. Wiersma, and
  M.~Inguscio, Phys. Rev. Lett. {\bf 95}, 170410 (2005).

\bibitem{Schulte2005}
T.~Schulte, S.~Drenkelforth, J.~Kruse, W.~Ertmer, J.~Arlt, K.~Sacha,
  J.~Zakrzewski, and M.~Lewenstein, Phys. Rev. Lett. {\bf 95}, 170411 (2005).

\bibitem{Clement2006}
D.~Cl\'{e}ment, A.~F. Var\'{o}n, J.~A. Retter, L.~Sanchez-Palencia, A.~Aspect,
  and P.~Bouyer, New J. Phys. {\bf 8}, 165 (2006).

\bibitem{Chen2007}
Y.~P. Chen, J.~Hitchcock, D.~Dries, M.~Junker, C.~Welford, and R.~G. Hulet,
Phys. Rev. A {\bf 77}, 033632 (2008).

\bibitem{Clement2008}
D.~Cl\'{e}ment, P.~Bouyer, A.~Aspect, and L.~Sanchez-Palencia, 
Phys. Rev. A {\bf 77}, 033631 (2008).

\bibitem{Paredes2005}
B.~Paredes, F.~Verstraete, and J.~I. Cirac, Phys. Rev. Lett. {\bf 95}, 140501
  (2005).

\bibitem{Gavish2005}
U.~Gavish and Y.~Castin, Phys. Rev. Lett. {\bf 95}, 020401 (2005).

\bibitem{Diener2001}
R.~B. Diener, G.~A. Georgakis, J.~Zhong, M.~Raizen, and Q.~Niu, Phys. Rev. A
  {\bf 64}, 033416 (2001).

\bibitem{Damski2003}
B.~Damski, J.~Zakrzewski, L.~Santos, P.~Zoller, and M.~Lewenstein, Phys. Rev.
  Lett. {\bf 91}, 080403 (2003).

\bibitem{Lye2007}
J.~E. Lye, L.~Fallani, C.~Fort, V.~Guarrera, M.~Modugno, D.~S. Wiersma, and
  M.~Inguscio, Phys. Rev. A {\bf 75}, 061603(R) (2007).

\bibitem{Fallani2007}
L.~Fallani, J.~E. Lye, V.~Guarrera, C.~Fort, and M.~Inguscio, Phys. Rev. Lett.
  {\bf 98}, 130404 (2007).

\bibitem{Guarrera2007}
V.~Guarrera, L.~Fallani, J.~E. Lye, C.~Fort, and M.~Inguscio, New J. Phys. {\bf
  9}, 107 (2007).

\bibitem{Guarrera2008}
V.~Guarrera, N.~Fabbri, L.~Fallani, C.~Fort, K.~M.~R. {van der Stam}, and
  M.~Inguscio, arXiv:0803.2015.

\bibitem{Aubry1980}
S.~Aubry and G.~André, Ann. Israel Phys. Soc {\bf 3}, 133 (1980).

\bibitem{Simon1982} 
B.~Simon, Adv. Appl. Math {\bf 3}, 463 (1982); J.~B. Sokoloff, Phys.
Rep. {\bf 126}, 189 (1985); H.~Hiramoto and M.~Kohmoto, Int. J. of
Mod. Phys. B {\bf 6}, 281 (1992).

\bibitem{Thouless1983}
D.~J. Thouless, Phys. Rev. B {\bf 28}, 4272 (1983).

\bibitem{Kohmoto1983}
M.~Kohmoto, L.~P. Kadanoff, and C.~Tang, Phys. Rev. Lett. {\bf 50}, 1870
  (1983).

\bibitem{Kohmoto1983a}
M.~Kohmoto, Phys. Rev. Lett. {\bf 51}, 1198 (1983);
C.~Tang and M.~Kohmoto, Phys. Rev. B {\bf 34}, 2041 (1986).

\bibitem{Zhong1995}
J.~X. Zhong and R.~Mosseri, J. Phys.: Cond. Matt. {\bf 7}, 8383 (1995);
F.~Pi\'echon, Phys. Rev. Lett. {\bf 76}, 4372 (1996).

\bibitem{Vidal1999}
J.~Vidal, D.~Mouhanna, and T.~Giamarchi, Phys. Rev. Lett. {\bf 83}, 3908
  (1999); Phys. Rev. B {\bf 65}, 014201 (2001).

\bibitem{Roth2003}
R.~Roth and K.~Burnett, Phys. Rev. A {\bf 68}, 023604 (2003).

\bibitem{Bar-Gill2006}
N.~Bar-Gill, R.~Pugatch, E.~Rowen, N.~Katz, and N.~Davidson,
  arXiv:cond-mat/0603513.

\bibitem{Louis2006}
P.~J.~Y. Louis and M.~Tsubota, arXiv:cond-mat/0609195.

\bibitem{Buonsante2004}
P.~Buonsante and A.~Vezzani, Phys. Rev. A {\bf 70}, 033608 (2004).

\bibitem{Buonsante2005}
P.~Buonsante, V.~Penna, and A.~Vezzani, Phys. Rev. A {\bf 72}, 031602(R) (2005).

\bibitem{Rousseau2006}
V.~G. Rousseau, D.~P. Arovas, M.~Rigol, F.~H\'{e}bert, G.~G. Batrouni, and
  R.~T. Scalettar, Phys. Rev. B {\bf 73}, 174516 (2006).

\bibitem{Roscilde2007}
T.~Roscilde, Phys. Rev. A {\bf 77}, 063605 (2008).

\bibitem{wikicontinuedfraction}
http://en.wikipedia.org/wiki/Continued\_fraction.

\bibitem{Haldane1981}
F.~D.~M. Haldane, Phys. Rev. Lett. {\bf 47}, 1840 (1981).

\bibitem{Cazalilla2004}
M.~A. Cazalilla, J. Phys. B {\bf 37}, S1 (2004).

\bibitem{Giamarchi2004}
T.~Giamarchi, {\em Quantum Physics in one Dimension} International series of
  monographs on physics Vol. 121 (Oxford University Press, Oxford, UK, 2004).

\bibitem{Dzhaparidze1978}
G.~I. Dzhaparidze and A.~A. Nersesyan, JETP Lett. {\bf 27}, 334 (1978);
V.~L. Pokrovsky and A.~L. Talapov, Phys. Rev. Lett. {\bf 42}, 65 (1979);
H.~J. Schulz, Phys. Rev. B {\bf 22}, 5274 (1980).

\bibitem{Kosterlitz1973}
J.~M. Kosterlitz and D.~J. D~J~Thouless, J. Phys. C: Solid State Phys. {\bf 6},
  1181 (1973); J.~M. Kosterlitz, J. Phys. C: Solid State Phys. {\bf 7}, 1046 (1974);
V.~L. Berezinskii, Sov. Phys. JETP {\bf 32}, 493 (1971).

\bibitem{Giamarchi1997}
T.~Giamarchi, Physica B {\bf 230}, 975 (1997).

\bibitem{Kuhner2000}
T.~D. K\"uhner, S.~R. White, and H.~Monien, Phys. Rev. B {\bf 61}, 12474
  (2000).

\bibitem{Arlego2001}
M.~Arlego, D.~C. Cabra, and M.~D. Grynberg, Phys. Rev. B {\bf 64}, 134419
  (2001).

\bibitem{Rigol2005}
M.~Rigol and A.~Muramatsu, Mod. Phys. Lett. B {\bf 19}, 861 (2005).

\bibitem{White1992}
S.~R. White, Phys. Rev. Lett. {\bf 69}, 2863 (1992).

\bibitem{White1993}
S.~R. White, Phys. Rev. B {\bf 48}, 10345 (1993).

\bibitem{Schollwoeck2005}
U.~Schollwöck, Rev. Mod. Phys. {\bf 77}, 259 (2005).

\bibitem{Schmitteckert1998}
P.~Schmitteckert, T.~Schulze, C.~Schuster, P.~Schwab, and U.~Eckern, Phys. Rev.
  Lett. {\bf 80}, 560 (1998).

\bibitem{Hida2001}
K.~Hida, Phys. Rev. Lett. {\bf 86}, 1331 (2001).

\bibitem{Schuster2002}
C.~Schuster, R.~A. R\"omer, and M.~Schreiber, Phys. Rev. B {\bf 65}, 115114
  (2002).

\bibitem{McCulloch2007}
I.~P. McCulloch, J. Stat. Mech.: Theor. Exp. , P10014 (2007).

\bibitem{Barache1994}
D.~Barache and J.~M. Luck, Phys. Rev. B {\bf 49}, 15004 (1994).

\bibitem{Kollath2004}
C.~Kollath, U.~Schollw\"{o}ck, J.~von Delft, and W.~Zwerger, Phys. Rev. A {\bf
  69}, 031601(R) (2004).

\bibitem{Altman2004}
E.~Altman, E.~Demler, and M.~D. Lukin, Phys. Rev. A {\bf 70}, 013603 (2004).

\bibitem{Rey2006}
A.~M. Rey, I.~I. Satija, and C.~W. Clark, Phys. Rev. A {\bf 73}, 063610 (2006).

\bibitem{Rigol2004a}
M.~Rigol and A.~Muramatsu, Phys. Rev. Lett. {\bf 93}, 230404 (2004).

\bibitem{Sanchez-Palencia2007}
L.~Sanchez-Palencia, D.~Cl\'{e}ment, P.~Lugan, P.~Bouyer, G.~V. Shlyapnikov,
  and A.~Aspect, Phys. Rev. Lett. {\bf 98}, 210401 (2007);
B.~Shapiro, Phys. Rev. Lett. {\bf 99}, 060602 (2007).

\bibitem{Rodriguez2006}
K.~Rodriguez, S.~R. Manmana, M.~Rigol, R.~M. Noack, and A.~Muramatsu, New J.
  Phys. {\bf 8}, 169 (2006).

\bibitem{Trombettoni2001}
A. Trombettoni and A. Smerzi, Phys. Rev. Lett {\bf 86}, 2353 (2001).

\bibitem{Anker2005} 
Th.~Anker, M.~Albiez, R.~Gati, S.~Hunsmann, B.~Eiermann, A.~Trombettoni, 
and M.~K. Oberthaler, Phys. Rev. Lett {\bf 94}, 020403 (2005).


\end{thebibliography}
\end{document}